\documentclass[pra,showpacs,superscriptaddress,amssymb,twocolumn,longbibliography,nofootinbib]{revtex4-1}

\usepackage{float}
\usepackage{amsmath,amssymb,setspace}
\usepackage{graphicx}
\usepackage{subfigure}  
\usepackage[usenames,dvipsnames]{color}
\usepackage{graphicx}
\usepackage{bm, bbm}      
\usepackage{color}
\usepackage{srcltx}
\usepackage{hyperref}
\hypersetup{
    colorlinks=true,       
    linkcolor=cyan,          
    citecolor=magenta,        
    urlcolor=cyan,           
    runcolor=cyan
}
\usepackage{upgreek}
\usepackage[normalem]{ulem}
\usepackage{slashed}

\usepackage[utf8]{inputenc}
\usepackage[english]{babel}

\def\sqr#1#2{{\vcenter{\vbox{\hrule height.#2pt \hbox{\vrule width.#2pt height#1pt \kern#1pt \vrule width.#2pt}\hrule height.#2pt}}}}

\def\beq#1{\begin{equation} \label{#1}}
\def\eeq{\end{equation}}
\def\ben{\begin{equation*}}
\def\een{\end{equation*}}
\def\bequa{\begin{eqnarray}}
\def\eequa{\end{eqnarray}}

\def\Tr{\mathop{\mathrm{Tr}}}

\def\sgn{\mathop{\mathrm{sgn}}}
\def\tr{\text{tr}}

\def\six{S_i^x}
\def\siz{S_i^z}
\def\sixt{\tilde{S}_i^x}
\def\sizt{\tilde{S}_i^z}
\def\sta{\sin \theta}
\def\cta{\cos \theta}
\def\aidag{a_i^{\dagger}}
\def\b0d{b_0^{\dagger}}
\def\bkd{b_k^{\dagger}}
\def\bmkd{b_{-k}^{\dagger}}

\def\sumk{\sum_k}
\def\sumi{\sum_{i=1}^N}

\def\tr{\text{tr}}

\newcommand{\ignore}[1]{}

\newcommand\bea{\begin{eqnarray}}
\newcommand\eea{\end{eqnarray}}
\newcommand{\bes}{\begin{subequations}}
\newcommand{\ees}{\end{subequations}}

\def\beq{\begin{equation}}
\def\eeq{\end{equation}}

\def\al{\alpha}

\newcommand{\be}{\begin{equation}}
\newcommand{\ee}{\end{equation}}
\newcommand{\ba}{\begin{align}}
\newcommand{\ea}{\end{align}}
\newcommand{\bg}{\begin{gather}}
\newcommand{\eg}{\end{gather}}
\newcommand{\bseq}{\begin{subequations}}
\newcommand{\eseq}{\end{subequations}}

\begin{document}

\title{Nested Quantum Annealing Correction at Finite Temperature: $p$-spin models}

\author{Shunji Matsuura}
\affiliation{1QB Information Technologies (1QBit), Vancouver, British Columbia, V6C 2B5, Canada}
\affiliation{Niels Bohr International Academy and Center for Quantum Devices,
Niels Bohr Institute, Copenhagen University, Blegdamsvej 17, Copenhagen, Denmark}
\author{Hidetoshi Nishimori}
\affiliation{Department of Physics, Tokyo Institute of Technology, Oh-okayama, Meguro-ku, Tokyo 152-8551, Japan}
\author{Walter Vinci}
\affiliation{Department of Physics and Astronomy, University of Southern California, Los Angeles, California 90089, USA}
\affiliation{Department of Electrical Engineering, University of Southern California, Los Angeles, California 90089, USA}
\affiliation{Center for Quantum Information Science \& Technology, University of Southern California, Los Angeles, California 90089, USA}
\author{Daniel A. Lidar}
\affiliation{Department of Physics and Astronomy, University of Southern California, Los Angeles, California 90089, USA}
\affiliation{Center for Quantum Information Science \& Technology, University of Southern California, Los Angeles, California 90089, USA}
\affiliation{Department of Electrical Engineering, University of Southern California, Los Angeles, California 90089, USA}
\affiliation{Department of Chemistry, University of Southern California, Los Angeles, California 90089, USA}

\begin{abstract}
Quantum annealing in a real device is necessarily susceptible to errors due to diabatic transitions and thermal noise.
Nested quantum annealing correction is a method to suppress errors by using an all-to-all penalty coupling among a set of physical qubits representing a logical qubit. We show analytically that nested quantum annealing correction can suppress errors effectively in ferromagnetic and antiferromagnetic Ising models with infinite-range interactions.
Our analysis reveals that the nesting structure can significantly weaken or even remove first-order phase transitions, in which the energy gap closes exponentially.
The nesting structure also suppresses thermal fluctuations by reducing the effective temperature.
\end{abstract} 
\maketitle

\section{Introduction}

There has been much interest in developing methods based on quantum mechanics to solve classically hard problems.
Among such methods, quantum annealing (QA)  emerged as a quantum metaheuristic to solve hard optimization problems~\cite{finnila_quantum_1994,kadowaki_quantum_1998,Farhi:01,Brooke1999,brooke_tunable_2001,Santoro,RevModPhys.80.1061,Farhi:01}.
The algorithm attempts to find a configuration of variables that minimize a cost function, or a problem Hamiltonian, by carefully exploiting the effects of quantum fluctuations. As such, a physical implementation of the algorithm in quantum hardware is of particular interest, since it might lead to quantum speedups over algorithms running on classical hardware. Following pioneering QA experiments in naturally occurring disordered magnets~\cite{Brooke1999,brooke_tunable_2001}, theoretical proposals~\cite{2002quant.ph.11152K,Kaminsky-Lloyd} inspired the construction of commercial, programmable quantum annealing processors using superconducting flux qubits by D-Wave Systems Inc.~\cite{harris_flux_qubit_2010,Harris:2010kx,Berkley:2010zr,Dwave,Berkley:2013bf,Bunyk:2014hb,DWave-entanglement,DWave-16q}. These processors implement the transverse field Ising model and have been the subject of intense independent scrutiny~\cite{q-sig,q108,Albash:2014if,Albash:2015pd,Smolin,comment-SS,SSSV,SSSV-comment}, as the search for examples of quantum speedup using them continues~\cite{speedup,Boixo:2014yu,Venturelli:2014nx,DW2000Q,McGeoch,King:2015cs,Hen:2015rt,Katzgraber:2015gf,McGeoch,PhysRevX.6.031015,2016arXiv160401746M,King:2015cs,Vinci:2016tg,Albash:2017aa,Mandra:2017ab}. They have also inspired alternative approaches to QA with more coherent but far less numerous flux qubits~\cite{Yan:2016aa,Weber:2017aa}. Alternative approaches that implement the transverse field Ising model using atomic systems such as ion traps \cite{Islam:2013mi,Smith:2016aa}, quantum gas microscopes~\cite{Simon:2011aa,Boll:2016aa}, and Rydberg atoms \cite{Weimer:2010aa,Nguyen:2017aa} are also being actively pursued, primarily for the purpose of quantum simulation of interacting quantum spin systems, but such systems are likely to be flexible enough to also be used eventually for solving combinatorial optimization problems, and have already broken the $50$ qubit barrier \cite{Bernien:2017aa,Zhang:2017aa}.

The idea of using quantum hardware to realize algorithmic speedups of course has deep roots in quantum computation~\cite{Feynman:1985ul,Feynman1}, which provided the first theoretical examples of quantum speedups~\cite{Deutsch:92,Bernstein:93,Shor:94,Grover:1996}. Adiabatic quantum computing (AQC), or the quantum adiabatic algorithm~\cite{Farhi:00,Farhi:01}, is a QA-inspired idea that formalizes the requirements for a quantum speedup using the quantum adiabatic theorem~\cite{Born:1928,Kato:50}. This theorem (or set of theorems~\cite{Garrido:62,nenciu_adiabatic_1980,Avron:87,Hagedorn:02,Jansen:07,lidar:102106,Ge:2015wo,Bachmann:2017aa})
guarantees that the minimum of a problem Hamiltonian can be found by an adiabatic process in a closed system, provided the computational time grows inversely with a small power (typically $2$ or $3$) of the minimum gap encountered during the adiabatic evolution. If this gap closes more slowly with the  problem size than the best classical algorithm for the same problem, then the adiabatic algorithm provides a quantum speedup; otherwise it may result in a slowdown (the adiabatic theorem only provides a sufficient condition). Examples of both types of scenarios are known and have been reviewed in Ref.~\cite{Albash-Lidar:RMP}. 

Just like any other quantum information processing method, the performance of real quantum annealers is hindered by decoherence and control errors. The former are due to the presence of the omnipresent (thermal) bath interacting with the physical device, and the latter are due to the analog nature of the computation. Therefore, the development of error correction techniques is crucial for the scalability and ultimate usefulness of QA hardware.
Various error suppresion methods for AQC have been proposed and studied theoretically~\cite{jordan2006error,PhysRevA.86.042333,Young:13,Sarovar:2013kx,Young:2013fk,Bookatz:2014uq,Marvian:2014nr,Jiang:2015kx,Marvian:2016kb,Marvian-Lidar:16,PhysRevLett.100.160506,Ganti:13}. An error correction approach that can be directly implemented on the current generations of quantum annealers, called quantum annealing correction (QAC), has been developed and tested experimentally~\cite{PAL:13,PAL:14,Mishra:2015,Vinci:2015jt,vinci2015nested,Vinci:2017ab}.

In QAC, logical qubits are redundantly encoded into a larger number of physical qubits. Energy penalties which commute with the problem Hamiltonian but anticommute with undesired spin-flips occurring late in the computation (errors) are then introduced in order to induce the physical qubits representing each logical variable to represent it consistently throughout the anneal. At the end of the anneal a decoding procedure is performed, which enables the recovery of logical ground-state candidates even if errors did accumulate during the anneal and the final state is not in the code space. Three types of QAC have been introduced and studied. The first uses designated penalty qubits to impose the energy penalty~\cite{PAL:13,PAL:14,Mishra:2015} [which we refer to as ``penalty QAC" (PQAC)],  while the other two use no designated penalty qubits but connect physical qubits to impose penalties against excitations~\cite{Vinci:2015jt,vinci2015nested,Vinci:2017ab}. Of these, one uses a two-level grid structure to connect physical qubits~\cite{Vinci:2015jt}, while the other uses a nesting of complete graphs for this purpose. Here we are primarily interested in the latter, known as ``nested QAC" (NQAC). Experiments using four different generations of the D-Wave quantum annealing devices show that all three methods increase the success probability of QA significantly~\cite{PAL:13,PAL:14,Vinci:2015jt,Mishra:2015,vinci2015nested,Vinci:2017ab}. In one case the scaling of the time-to-solution in the case of random Ising problems was shown to improve using PQAC~\cite{PAL:14}, which to date is the only demonstration of error correction improving the \emph{algorithmic} performance of a quantum information processing device.
 
Previously, we used a mean field analysis to study the performance of PQAC for the transverse field Ising model in both the ferromagnetic and Hopfield model cases, the latter involving randomness and frustration~\cite{MNAL:15,Matsuura:2016aa}. We showed analytically that PQAC can prevent or `soften' a quantum phase transition, depending on its order. 
In PQAC, the designated penalty qubits behave as a source of external fields. At zero temperature, this makes the paramagnetic phase unstable and 
induces symmetry breaking from the paramagnetic phase to the ferromagnetic or Hopfield phase.
At finite temperature, the paramagnetic phase is not destabilized by the penalty term. Nevertheless, the effective free energy barrier between the two phases becomes significantly smaller as the penalty coupling increases, thus enhancing tunneling.

Here, we use the same approach to analytically study NQAC at zero and finite temperatures. Because of the high connectivity in NQAC, the mean field approach is particularly suitable.

The remainder of this paper is organized as follows.
We first review NQAC in Sec.~\ref{Nested quantum annealing}. 
In Sec.~\ref{Nested Quantum Annealing correction in ferromagnetic system}, we study the phase diagrams of ferromagnetic systems after NQAC by analyzing the free energy and saddle point equations. We show how NQAC changes the properties of phase transitions depending on the type of interactions.
In Sec.~\ref{antiferromagnetic case}, we study the antiferromagnetic case.
Section~\ref{Penalty term in the nested QAC} discusses the effect of introducing designated penalty qubits in NQAC. We conclude with a discussion in Sec.~\ref{sec:discussion}.
Details of various calculations and additional results are given in the Appendix.

\section{Hamiltonian for nested quantum annealing correction} 
\label{Nested quantum annealing}

Many combinatorial optimization problems can be formulated in terms of a classical Ising Hamiltonian~\cite{2013arXiv1302.5843L},
\bea
H^{Z}=-\sum_{i}h_{i}\sigma^{z}_{i}-\sum_{(i,j)}J_{ij}\sigma^{z}_{i}\sigma^{z}_{j}\ ,
\label{prob-hamil}
\eea
where $i=1,2,\dots ,N$ are site as well as qubit indices, $\{h_i\}$ the local fields, $\{J_{ij}\}$ the Ising couplings, and $\sigma^{\alpha}_{i}$ is the $\alpha\in\{x,y,z\}$-component of the Pauli matrix acting on the $i$-th qubit.
The couplings $\{h_i\}$ and $\{J_{ij}\}$ are chosen so that the ground state of $H^{Z}$ represents the solution to the optimization problem we wish to solve.
Quantum fluctuations are induced by a driver Hamiltonian.
We consider the standard form,
\bea
H^{X}=-\sum_{i=1}^{N}\sigma^{x}_{i}\ ,
\label{Hx}
\eea
whose ground state is easy to prepare, and which has a small but non-vanishing overlap with any eigenstate of $H^{Z}$.
The time-dependent Hamiltonian is given by
\begin{align}
\label{eq:Hdef}
H=A(t) H^{X}+ B(t)H^{Z} \equiv B(t)\left(\Gamma(t) H^{X}+H^{Z}\right)\ .
\end{align}
The initial state at $t = 0$ is the ground state of the driver Hamiltonian [$A(0)=1, B(0)=0$].  We adiabatically reduce quantum fluctuations to reach the problem Hamiltonian at the final time $t=t_f$ [$A(t_f)=0, B(t_f)=1$].

Fluctuations induced by the driver term are desirable but are deliberately made less likely towards the end of the anneal by decreasing the magnitude of the driver term. However, undesirable bath-induced fluctuations can cause bit-flip errors at all times, including when $A(t) \ll B(t)$, which generally result in transitions out of the ground state. To reduce the likelihood of such undesirable fluctuations we use  NQAC to encode the Hamiltonian~\cite{vinci2015nested}. The basic idea is to (1) increase the energy cost of an undesirable bit-flip by creating ferromagnetically coupled logical qubits, (2) decode any bit-flip errors that might have nevertheless taken place. 

We label a physical qubit within the $i$-th logical qubit by $(i,c)$, with the logical qubit index $i=1,\cdots, N$ and the nesting index $c=1,\cdots, C$. The physical problem Hamiltonian~\eqref{prob-hamil} is first promoted to a logical problem Hamiltonian, whose Pauli matrices $\sigma^{z}_{i}$ represent the logical binary ($\pm 1$) variables of the given optimization problem.
All indices in the logical problem Hamiltonian
are then mapped to physical qubit indices as $\sigma^{z}_{i}\mapsto \sigma^{z}_{ic}$, $J_{ij}\mapsto J_{(i,c)(j,c')}$, and $h_{i}\mapsto h_{(i,c)}$. 
Logical qubits $i$ and $j$ interact with strength $J_{ij}$. An obvious way to realize this is to connect all physical qubits $(i,c)$ and $(j,c')$ with the same strength, and correspondingly rescale the local fields:
\bes
\label{interactions}
\begin{align}
J_{(i,c),(j,c')}&=J_{ij},~~~ \forall c,c'~~~\forall i\neq j\\
h_{(i,c)}&=C h_i,~~~ \forall c,i \ .
\end{align}
\ees
The need to rescale the local fields by $C$ can be viewed to some extent as a deficiency of NQAC since it requires growing the energy scale; it can of course be avoided by restricting ourselves to problem instances without local fields.  Reducing the energy scale by rescaling $J$ to $J/C$ is not necessarily preferable, since in practice, with finite precision couplers as is the case with the current generation of commercial quantum annealers, this means a loss of precision.

Next is the penalty term. In NQAC we couple all $C$ physical qubits of logical qubit $i$ ferromagnetically with strength $\gamma$, 
\beq
J_{(i,c),(i,c')}=\gamma,~~~ \forall c\neq c' ~~~ \forall i\ ,
\eeq
so that bit-flip errors are suppressed by an energy gap. As long as no more than $\lfloor (C-1)/2\rfloor$ of the qubits have flipped, we can faithfully decode the logical qubit. In other words, we can equivalently view this procedure as encoding a logical qubit into a distance-$C$ repetition code. The $N$ penalty terms (one for each logical qubit) 
\begin{align}
H_{\mathrm{pen},i} = -\gamma\left(\sum_{c_i=1}^{C} \sigma_{ic_i}^{z}\right)^{2} = -\gamma \left[C I + 2 \sum_{c<c'}^C \sigma_{ic}^{z}\sigma_{ic'}^{z}\right]
\end{align}
are added to the total Hamiltonian. Note that this sum includes $C$ identity terms which merely shift the total energy, and that we are double-counting the interactions; the latter means that the physical penalty strength is actually $2\gamma$. Note further that each $\sigma_{ic}^{z}\sigma_{ic'}^{z}$ term can be understood as an element of the stabilizer of the distance-$C$ repetition code; because of the complete graph ($K_C$) connectivity of the physical qubits, each single-qubit bit-flips error $\sigma^x_{ic}$ anticommutes with (is detected by) $C-1$ such terms, which is another way to understand the energy penalty using the ideas of error suppression for AQC~\cite{jordan2006error,PhysRevA.86.042333,Young:13,Sarovar:2013kx,Young:2013fk,Bookatz:2014uq,Marvian:2014nr,Jiang:2015kx,Marvian:2016kb,Marvian-Lidar:16}.

Since current QA devices do not extend control to the driver Hamiltonian, which is fixed as a simple sum over single-qubit $\sigma^x$ terms, the new driver Hamiltonian replacing Eq.~\eqref{Hx} is simply 
\beq
H^{X}=-\sum_{i=1}^{N}\sum_{c_i=1}^C\sigma^{x}_{ic_i}\ .
\label{eq:newHx}
\eeq
Note that this means that the driver Hamiltonian appears as an `error' detected by the penalty Hamiltonian. This is clearly an undesirable aspect of NQAC, but as we shall see it will fortunately not completely spoil the method's error suppression ability. Intuitively, this can be understood as being due to the fact that the suppression effect is small when fluctuations are deliberate [i.e., when $B(t) \ll A(t)$]. Conversely, the suppression effect peaks when the deliberate fluctuations are small [i.e., when $B(t) \gg A(t)$], and this is exactly when undesirable fluctuations would be potent if left unsuppressed. The potentially problematic regime is when $B(t) \approx A(t)$; our analysis below reveals that the problem is in fact not severe. Finally, note that to also encode the driver Hamiltonian using the logical operators of the repetition code (as in the AQC schemes~\cite{jordan2006error,PhysRevA.86.042333,Young:13,Sarovar:2013kx,Young:2013fk,Bookatz:2014uq,Marvian:2014nr,Jiang:2015kx,Marvian:2016kb,Marvian-Lidar:16}) would require mapping $\sigma^{x}_{i} \mapsto \otimes_{c=1}^C\sigma^{x}_{ic}$, which is unfeasible.

\section{Second-order transition in the ferromagnetic model}
\label{Nested Quantum Annealing correction in ferromagnetic system}

We first consider a mean-field analysis of an NQAC implementation of a fully ferromagnetic system with two-body interactions. As we shall see, this system undergoes a second order phase transition during the anneal. Our analysis in this section reproduces in more detail some of the results first reported in Ref.~\cite{vinci2015nested}.

\subsection{Problem formulation and the free energy}

We choose the logical problem Hamiltonian to be the infinite-range ferromagnetic Ising model with two-body interactions,
\bea
H^{Z}=-\frac{J}{N}\left(\sum_{i}\sigma^{z}_{i}\right)^{2}.
\eea
The encoded Hamiltonian takes the following form according to Eq.~\eqref{interactions},
\bes
\begin{align}
\label{ferro-hamil}
H=
&-\Gamma H^{X} -{J\over N}\sum_{i\neq j}^{N}\sum_{c_i,c_j=1}^{C}  \sigma_{ic_i}^{z}\sigma_{jc_j}^{z}
-\sum_{i=1}^{N}H_{\mathrm{pen},i} \\
=&-\Gamma H^{X}-{NJ} \left({1\over N}\sum_{i=1}^{N}\sum_{c_i=1}^{C}\sigma_{ic_i}^{z}\right)^{2}
\notag \\
&-\lambda\sum_{i=1}^{N}\left(\sum_{c_i=1}^{C} \sigma_{ic_i}^{z}\right)^{2}
\label{ferro-hamil 0} 
\end{align}
\ees
where 
\beq
\lambda\equiv \gamma-{J/N}
\label{eq:lambda-def}
\eeq 
and from now on $H^X$ is as given in Eq.~\eqref{eq:newHx}. We are interested in the large $N$ limit in this work while keeping $J$ fixed, so that $\gamma\to \lambda$. We shall thus refer to $\lambda$ as the effective penalty strength. Note that in Eq.~\eqref{ferro-hamil} we have rescaled the logical couplings, as is customary in mean-field analyses, to ensure that the total Hamiltonian is extensive in $N$.
Our goal is to understand the free-energy landscape and the effects of the nesting level $C$ from the point of view of mean field theory.

For a given temperature $T=1/\beta$ and a Hamiltonian $H$, the partition function is given by $Z=\tr \, e^{-\beta H}$ (we use units where $k_{\text B}= \hbar =1$).
The free energy per single logical qubit is defined by $F=-(T/N) \log Z$. We take the following standard steps. First, we apply the Suzuki-Trotter decomposition~\cite{Suzuki:1976rt}
to rewrite the partition function in path integral form.
Then we introduce Hubbard-Stratonovich fields
$m_i$ for each logical qubit  and $m$ for the total magnetization,
\bes
\begin{align}
 S_i^{z(x)}&=\frac{1}{C}\sum_{c=1}^C \sigma_{ic}^{z(x)},~~~   S^z=\frac{1}{N}\sumi \siz \\
  m_{i}&=\left\langle S_i^{z} \right\rangle,~~~m=\left\langle S^{z} \right\rangle={1\over N}\sum_{i=1}^{N}m_i,
  \label{order parameter def}
  \end{align}
\ees
where brackets denote the standard thermal average.
We make the static approximation, i.e., we take the fields $m_i$ and $m$ to be constant along the Euclidean time direction. This is known to give the exact solution in the case of without nesting ($C=1$)~\cite{Seoane2012}.
The path integral is dominated by its saddle point for $m$ and $m_i$ when $N$ and $C$ are large.
A practically interesting case is  $N \gg C>1$, which corresponds to exploring the effects of NQAC on a computational problem with a large number of logical variables. We expect the mean field approximation within each logical qubit to be reasonably good already for moderately large $C$. By following the steps outlined above (see Appendix~\ref{Appendix derivation} for a detailed derivation) we find that the free energy is given by
\begin{align}
\label{free energy eq}
F&=JC^2m^2+{\lambda C^2\over N}\sum_{i=1}^{N}m_{i}^2 + \\
&-{C\over \beta N}\sum_{i=1}^{N}\log \left[2\cosh\left(\beta\sqrt{4C^2(Jm+\lambda m_i)^2+\Gamma^2} \right)\right]\notag \ .
\end{align}
We are interested in extracting the effect of nesting on the physical parameters of the problem. In particular, we would like to determine how $C$ rescales the temperature and the transverse field strength. Note that the quantum effects are captured by the last term in this free energy expression, since it explicitly contains the transverse field through its dependence on $\Gamma$.

\subsubsection{The low temperature limit}

In the low temperature limit $\beta\to\infty$,
\bea
F&\to&
JC^2m^2+{\lambda C^2\over N}\sum_{i=1}^{N}m_{i}^2 +  \cr
&-&{1\over  N}\sum_{i=1}^{N}\sqrt{4(JC^2m+\lambda C^2 m_i)^2+(\Gamma C)^2}. \cr
&&
\label{free energy 1}
\eea
One can see that, in this low temperature limit, the effect of $C$ levels of nesting is equivalent to rescaling all the parameters appearing in the free energy $F$ as follows: 
\bea
\beta \to \beta,~~J\to JC^2,~~\lambda \to \lambda C^2,~~\Gamma \to \Gamma C\,,
\label{scaling 1}
\eea
which can be interpreted as a boost of the problem couplings by a factor of $C^2$ while the driver term is boosted by a factor of $C$ only. The fact that the driver term scales differently than the problem Hamiltonian is due to the aforementioned fact that, in QAC, the driver term is not encoded.  

Alternatively, since the partition function is $Z=e^{- N \beta F}$, we can interpret the effect of nesting as being equivalent to the following rescaling 
\bea
\beta \to \beta C^2~~J \to J,~~\lambda \to \lambda,~~\Gamma \to \Gamma /C\ .
\label{scaling 2}
\eea
This can be understood as an effective temperature reduction by a factor of $C^2$, although at the same time the driver term is reduced by a factor of $C$, which corresponds to a suppression of quantum fluctuations; this once again reflects the price we have to pay for not encoding the driver Hamiltonian. The problem can be alleviated by rescaling $\Gamma$ to $C\Gamma$ by either increasing $A(t)$ or decreasing $B(t)$ [recall from Eq.~\eqref{eq:Hdef} that $\Gamma(t) = A(t)/B(t)$], but this presents a problem similar to the one discussed in relation to the rescaling of the local field $h_i$ following Eq.~\eqref{interactions}.

The effective temperature reduction was one of the main conclusions first reported in Ref.~\cite{vinci2015nested}, and we shall expand on it here: it is possible to `trade' physical qubits in exchange for a lower effective temperature. This provides an important mechanism for reducing thermal excitations occurring in physical implementations of quantum annealing. The nested structure of NQAC takes advantage of a more complex encoding to achieve better error correction than PQAC, in which the free energy is only linearly proportional to the number of physical qubits per logical qubit (the parameter $C$ in NQAC)~\cite{Matsuura:2016aa}.

\subsubsection{Arbitrary temperature}

We can derive the saddle-point equation for $m_i$ directly from Eq.~\eqref{free energy eq} without  taking the low temperature limit. The result is:
\bea
m_{i}&=&
{2(Jm+\lambda m_i)\over \sqrt{4(Jm+\lambda m_i)^2+(\Gamma/C)^2} }   \cr
&\times&  \tanh\left[
(\beta C)\sqrt{4(Jm+\lambda m_i)^2+(\Gamma/C)^2}
\right]. 
\label{ferro-saddle-b}
\eea
This shows that the effect of nesting is equivalent to rescaling the temperature and the transverse field as follows:
\bea
\beta\to\beta C,~~\Gamma \to \Gamma/C.
\label{scaling 3}
\eea
This rescaling is general and dictates the scaling of the critical points on the phase diagram as a function of the nesting level $C$. We can interpret it as showing that 
NQAC suppresses both thermal and quantum fluctuations. 

Interestingly, Eq.~\eqref{scaling 3} gives a result that is different from what we found in the low temperature limit, where we had $\beta \to \beta C^2$. 
This shows that NQAC is more effective at low temperatures. 

How do we reconcile these different results? First, note that there is in fact no inconsistency, since in the low temperature limit the $\beta$ dependence in Eq.~\eqref{ferro-saddle-b} disappears, so the saddle point equation makes no statement about the scaling of the temperature in this limit. Moreover, the scaling of $J$, $\gamma$, and $\Lambda$ implied by Eqs.~\eqref{scaling 1} and~\eqref{scaling 2} agree with their respective scaling from the saddle point equation~\eqref{ferro-saddle-b}.

More generally, in a physical system, it is the free energy (or equivalently the partition function) that determines the probability of finding certain states. Thus, Eq.~\eqref{free energy eq} is to be understood as the principal relation. However, it does not appear to be possible to directly extract the scaling relations from it, which is why we must resort to either the low temperature limit or the saddle point approximation.

It is useful to clarify the validity of the two corresponding scaling relations.
The scaling in Eqs.~\eqref{scaling 1} and~\eqref{scaling 2} is the scaling of the free energy or the partition function at low temperature. This means that any physical transition rates, or the probability of finding certain states, should follow this scaling at low temperature. 
On the other hand,  Eq.~\eqref{scaling 3} is the scaling of the saddle point solution. While this is valid for any temperature, it is not valid away from 
the saddle point and therefore it is not the scaling of the transition rates. 
Neither of the scalings are the exact scaling of the free energy in Eq.~\eqref{free energy eq}.

\subsection{Second-order phase transition}
\label{sec:2PT}

As we assume $J>0$ and $\lambda>0$, all interactions are  ferromagnetic, and the configuration that minimizes the free energy has all $m_i$ equal: $m_i=m$ $\forall i$. The saddle point equation then becomes simply that of the standard mean-field Ising model in a transverse field.
The solutions of the saddle point equation~\eqref{ferro-saddle-b} with $m_i=m$ depend on the scaled temperature $\beta C$ and the scaled transverse field $\Gamma/C$.
At a large temperature or a large transverse field, thermal or quantum fluctuations becomes large, and the order parameter vanishes: $m_i=m=0$.
As the temperature and the transverse field decrease, there is a second order phase transition to the ordered phase with $m_i=m$, which is doubly degenerate.

\begin{figure}[t]
\includegraphics[width=0.45\textwidth]{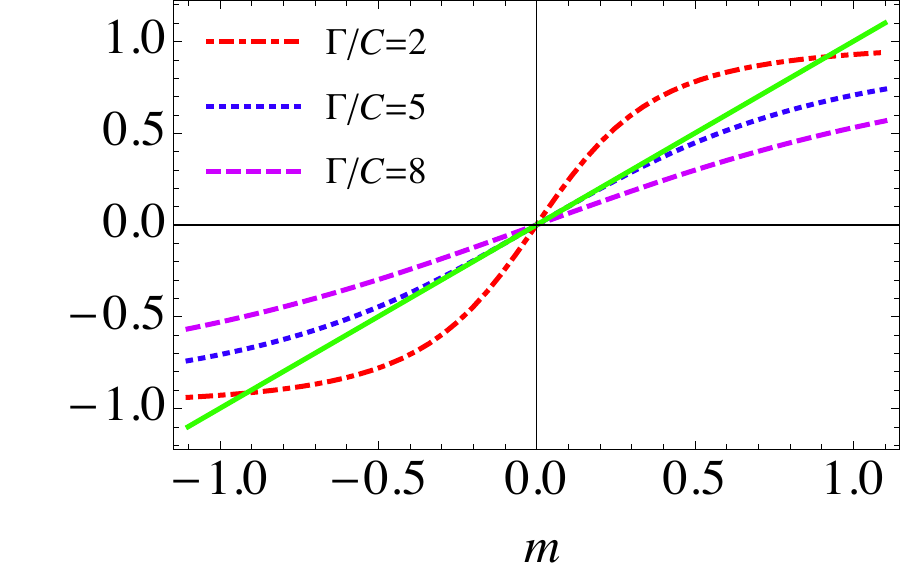} 
\caption{
Saddle point equation~\eqref{ferro-saddle-b} with $m=m_i$ at $J=1$, $\lambda=1.5$, $T/C=0.02$ with various values of $\Gamma$.
For $\Gamma/C<5$ there are three intersections (ferromagnetic phase), 
and above $\Gamma/C=5$ there is only one, $m=0$ (paramagnetic phase).}
\label{ground-eom}
\end{figure}

Figure~\ref{ground-eom} shows both sides of Eq.~\eqref{ferro-saddle-b} with $m_i=m$ for various values of $\Gamma/C$ at $J=1$, $\lambda=1.5$, $T/C=0.02$.
If $\Gamma/C$ is below the critical value $\Gamma\simeq5$, there are three intersections. At and above the critical point, the intersection is only at $m=0$, representing
the paramagnetic phase. Making $C$ larger at fixed $\Gamma$ moves the $m \neq 0$ intersections closer and closer to $m=\pm 1$, i.e., favors correct solution to the (in this case trivial) optimization problem. 

The critical values of $\beta C$ and $\Gamma/C$  satisfy
\bea
{\Gamma\over C}=2(J+\lambda)\tanh \left[\beta C  \left({\Gamma \over C}\right) \right].
\label{ferrosaddle}
\eea
At zero temperature, the critical $\Gamma$ is $\Gamma_{\rm c}=2C(J+\lambda)$, with the ferromagnetic (ordered) phase arising when $\Gamma < \Gamma_{\rm c}$.
This implies that both a larger penalty $\gamma$ and a larger nesting level $C$ favor the ordered phase. However, it is also clear that the role of the effective penalty is simply to shift the coupling $J$ to $J+\lambda$, which is a consequence of having set $m=m_i$. We will discuss this in more detail in Sec.~\ref{sec:FM for p and q}.

\begin{figure}[t]
\includegraphics[width=0.41\textwidth]{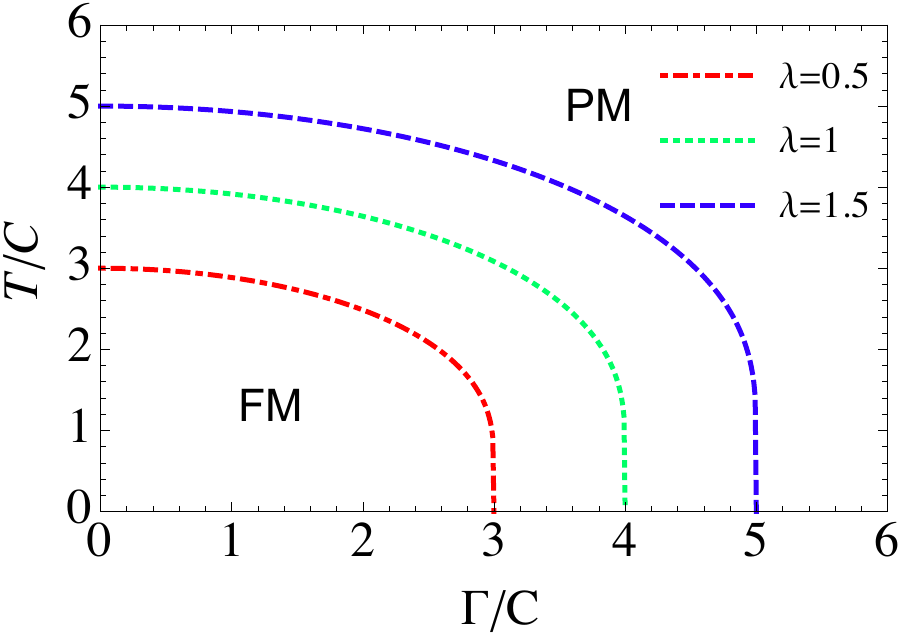} 
\caption{
The critical line in the $(\Gamma/C,T/C)$ plane for $J=1$ and $\lambda=0.5$, $1.0$ and $1.5$.
The ordered (disordered) phase is below (above) the lines.}
\label{groundgs}
\end{figure}

Clearly, the parameters in Eq.~\eqref{ferrosaddle} obey a scaling relation. If $\{\Gamma,\beta,(J+\lambda)\}$ is a solution, then
$\{\Gamma a,\beta/a,a(J+\lambda)\}$ is also a solution for arbitrary $a>0$. Thus it suffices to solve Eq.~\eqref{ferrosaddle} numerically for one set of values $\{\Gamma,\beta,(J+\lambda)\}$, and then rescale to get the rest.
The solution is shown in Fig.~\ref{groundgs} for various values of $\lambda$.
The ordered phase $m_i=m\ne 0$ is realized below each critical line, while the paramagnetic phase  
$m_i=m=0$ is realized above each line. Increasing the penalty favors the ordered phase (pushes the lines out).

\section{Ferromagnetic $p$-body interactions with $q$-body penalty}
\label{sec:FM for p and q}

Having studied the two-body case, we now move on to the generalization of NQAC to ferromagnetic systems with $p$-body interactions and $q$-body penalties. An NQAC Hamiltonian for the $p$-spin model with $q$-body penalty interactions within each logical qubit reads:
\begin{align}
\label{eq:H-general}
H=&-\Gamma\sum_{i}\sum_{c_i}\sigma^{x}_{ic_i}-JN\left({1\over N}\sum_{i}\sum_{c_i}\sigma^{z}_{ic_i} \right)^{p}
  \notag \\
&-\lambda \sum_{i}\left(\sum_{c_i} \sigma^{z}_{ic_i} \right)^{q}\ .
\end{align}
The first step is to determine the free energy and scaling relations, which we do next. The $p=q=2$ case we discussed in the previous section exhibits a second-order phase transition; as we shall see the generalized models undergo a first-order phase transitions for $p\ge 3$.

\subsection{Free energy and scaling relations}
\label{sec Problem_formulation}

By using the Suzuki-Trotter decomposition and the static ansatz, one can compute the free energy per logical qubit, to find:
\begin{align}
&F=(p-1)JC^p m^p+(q-1)\lambda C^{q} {1\over N}\sum_{i}m^q_{i}
-{C\over \beta N}\sum_{i} \times \notag \\
&\log \left[ 2\cosh \beta 
\left((pJ C^{p-1} m^{p-1}+q\lambda C^{q-1} m^{q-1}_i)^2+\Gamma^2\right)^{1/2}\right]
\label{general free energy eq}
\end{align}
with the same definitions of $m$ and $m_i$ as in Eq.~\eqref{order parameter def}.
The saddle-point equations are:
\bea
&&m=  {1\over N}\sum_{i}
{Jpm^{p-1}+\lambda C^{q-p}qm^{q-1}_{i} \over
\sqrt{({Jpm^{p-1}+\lambda C^{q-p}qm^{q-1}_{i})^2+(\Gamma/C^{p-1})^2} }} \cr
&&\times \tanh \beta C^{p-1}  \sqrt{({Jpm^{p-1}+\lambda C^{q-p}qm^{q-1}_{i})^2+(\Gamma/C^{p-1})^2} } \cr
&&m_i=
{Jpm^{p-1}+\lambda C^{q-p}qm^{q-1}_{i} \over
\sqrt{({Jpm^{p-1}+\lambda C^{q-p}qm^{q-1}_{i})^2+(\Gamma/C^{p-1})^2} }} \cr
&&\times \tanh \beta C^{p-1} \sqrt{({J pm^{p-1}+\lambda C^{q-p}qm^{q-1}_{i})^2+(\Gamma/C^{p-1})^2} }.  \cr
&&
\label{general ferro-saddle-b}
\eea
Each parameter scales with $C$ as
\bea
\lambda\to \lambda C^{q-p}, \beta\to\beta C^{p-1}, \Gamma\to\Gamma /C^{p-1}.
\label{scaling of parameters}
\eea
In the zero temperature limit, the free energy~\eqref{general free energy eq} reduces to the simple form
\bea
\label{free energy general pq}
F&=&(p-1)JC^p m^p+(q-1)\lambda C^{q} {1\over N}\sum_{i}m^q_{i} \\
&-&{C^{p}\over  N}
\sum_{i}  \sqrt{(pJ  m^{p-1}+q\lambda C^{q-p} m^{q-1}_i)^2+(\Gamma/C^{p-1})^2}. \notag
\eea
From this expression we see that the parameters appearing in the low-temperature free energy scale as:
\bea
\beta\to\beta, J\to J C^{p},\lambda\to \lambda C^{q}, \Gamma\to\Gamma  C\ ,
\label{scaling of parameters 2}
\eea
or, equivalently, using the partition function $Z=e^{- N \beta F}$:
\bea
\beta\to \beta C^{p}, J\to J ,\lambda\to \lambda C^{q-p}, \Gamma\to\Gamma C^{1-p}\ .
\eea
The same comments as in the previous section apply regarding the difference between the saddle point and low temperature results. An interesting new aspect of the present analysis is that the effective penalty strength $\lambda$ can be affected by the nesting level if $p\neq q$. Specifically, choosing $q<p$ causes $\lambda$ to decrease with increasing nesting after rescaling, as one might expect given that this choice lowers the degree of penalty interactions relative to the logical coupling between spins.

\subsection{First-order phase transition}
\label{First-order phase transition finite T}

\subsubsection{$p=q=4$}
\begin{figure}[t]
   \includegraphics[width=0.41\textwidth]{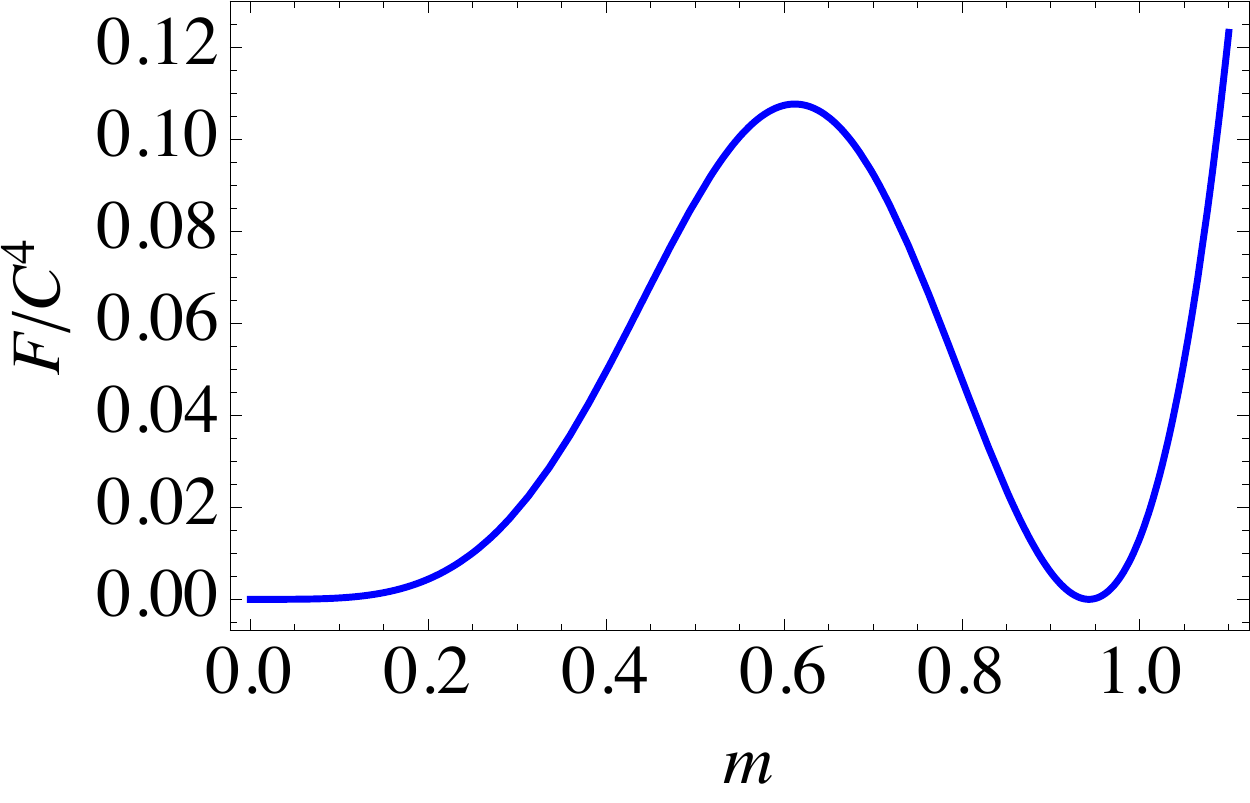}   
\caption{Free energy at the transition point normalized by $C^4$  as a function of $m$ for $p=q=4$. Parameters are chosen to be $J=1,\lambda=1,\Gamma/C^3=2.37,T/C^4=0.01.$  }
\label{freeb1j1G32}
\end{figure}

Let us begin the analysis of phase transitions with the case $p=q=4$. The free energy simplifies to
\bea
\label{free energy general p=q=4}
F&=&3C^4 \left(J m^4+\lambda  {1\over N}\sum_{i}m^4_{i} \right)\\
&-&2{C^{4}\over  N}
\sum_{i}  \sqrt{(J  m^3+\lambda m^{3}_i)^2+(\Gamma/C^{3})^2}. \notag
\eea
Near the end of  anneal, $\Gamma\sim0$, the free energy~\eqref{free energy general p=q=4} is proportional to $C^4$, and thus the effective temperature decreases as $T/C^4$.
If the temperature is very low one can expect a uniformly ordered state to be realized since the system is ferromagnetic. I.e., in this limit we expect $m=m_i$. The free energy at the transition point, exhibiting typical low temperature behavior ($T/C^4=0.01$), is plotted in Fig.~\ref{freeb1j1G32}, as a function of the magnetization $m$. The system clearly undergoes a first-order phase transition.
Our numerical results show that  the potential barrier at the first-order phase transition persists at any value of $\lambda$. This means that the potential barrier is not reduced by increasing $\lambda$, so that the penalty term is not helpful in this case. We can understand this as a consequence of the fact that when $p=q$, as can be seen from Eq.~\eqref{free energy general p=q=4}, the effective penalty $\lambda$ simply acts to shift $J$ (i.e., $J\to J+\lambda$) if $m_i = m$, as we also saw in the $p=q=2$ case discussed in Sec.~\ref{sec:2PT}.

We thus consider $p\neq q$, for which the role of the effective penalty $\lambda$ is different from the $p=q$ case, since then $\lambda$ cannot be absorbed into the ferromagnetic coupling $J$.

\subsubsection{$p=4$, $q=2$}

\begin{figure}[t]
   \includegraphics[width=0.45\textwidth]{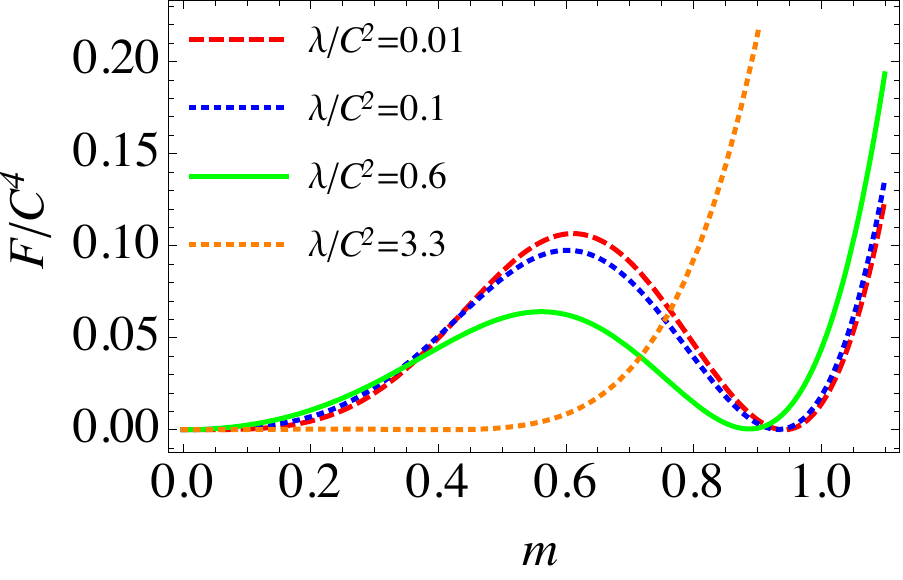}   
\caption{ 
Free energy at the transition point, normalized by $C^4$, as a function of $m$, for $p=4$, $q=2$, and $J=1,T/C^4=0.01.$    
The red dotted line is for $\lambda/C^2=0.01$ and $\Gamma/C^3=1.2$, the blue dashed line for $\lambda/C^2=0.1$ and $\Gamma/C^3=1.3$, the green solid line for $\lambda/C^2=0.6$ and $\Gamma/C^3=2.0$,
and the orange dotted line for $\lambda/C^2=3.3$ and $\Gamma/C^3=6.7$.}
\label{freec3b1nestedq4}
\end{figure}

The free energy at the transition point (where the free energy become degenerate) is plotted in Fig.~\ref{freec3b1nestedq4} for $p=4$ and $q=2$, for various values of $\lambda$.
One sees that the potential barrier becomes smaller as $\lambda$ increases, and it disappears for large values of $\lambda$. 


\begin{figure}[t]
\includegraphics[width=0.45\textwidth]{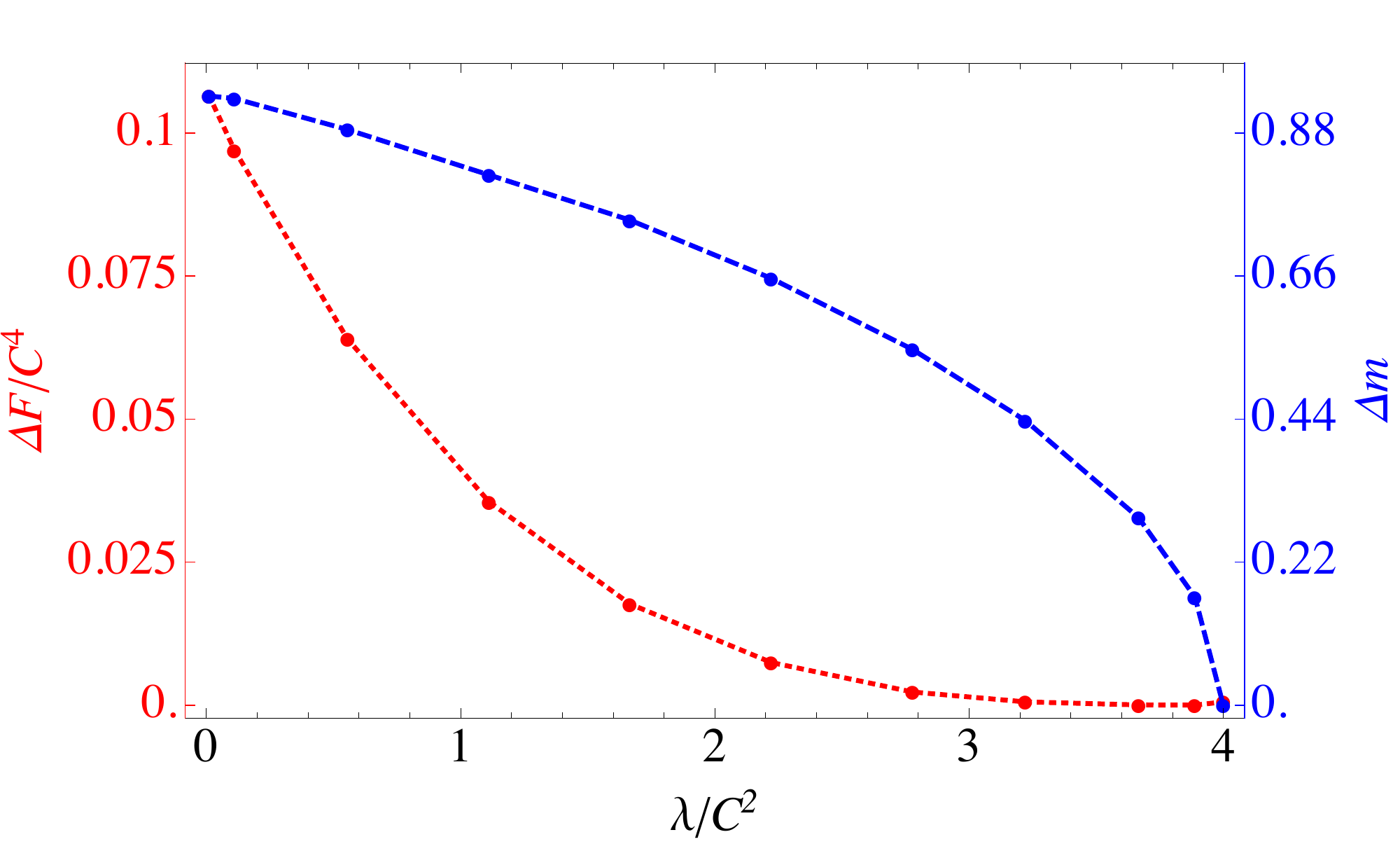}
\caption{The height of the barrier $\Delta F$ and the width $\Delta m$ at the transition point, for the same parameters as in Fig.~\ref{freec3b1nestedq4}. 
The dotted line (red) represents $\Delta F/C^4$ and  the dashed line (blue) represents $\Delta m$.
\label{fig:p=4}
}
\label{pgvsheight}
\end{figure}

The height ($\Delta F$) and the width ($\Delta m$) of the potential barrier at the transition point as functions of the effective penalty strength are shown in Fig.~\ref{pgvsheight}.
The system is in the paramagnetic phase for large $\Gamma \gg 1$ while it is in the ordered phase at $\Gamma=0$. Therefore, the phase transition exists for all values of $\lambda$. However, as $\lambda$ increases, the first-order phase transition in the ground state softens, in the sense of smaller values of the potential barrier height $\Delta F$ and width $\Delta m$, as seen in Fig.~\ref{pgvsheight}. The order of the phase transition eventually changes from first to second when the free energy barrier vanishes at $\lambda/C^2=4$.

\begin{figure}[t]
   \includegraphics[width=0.45\textwidth]{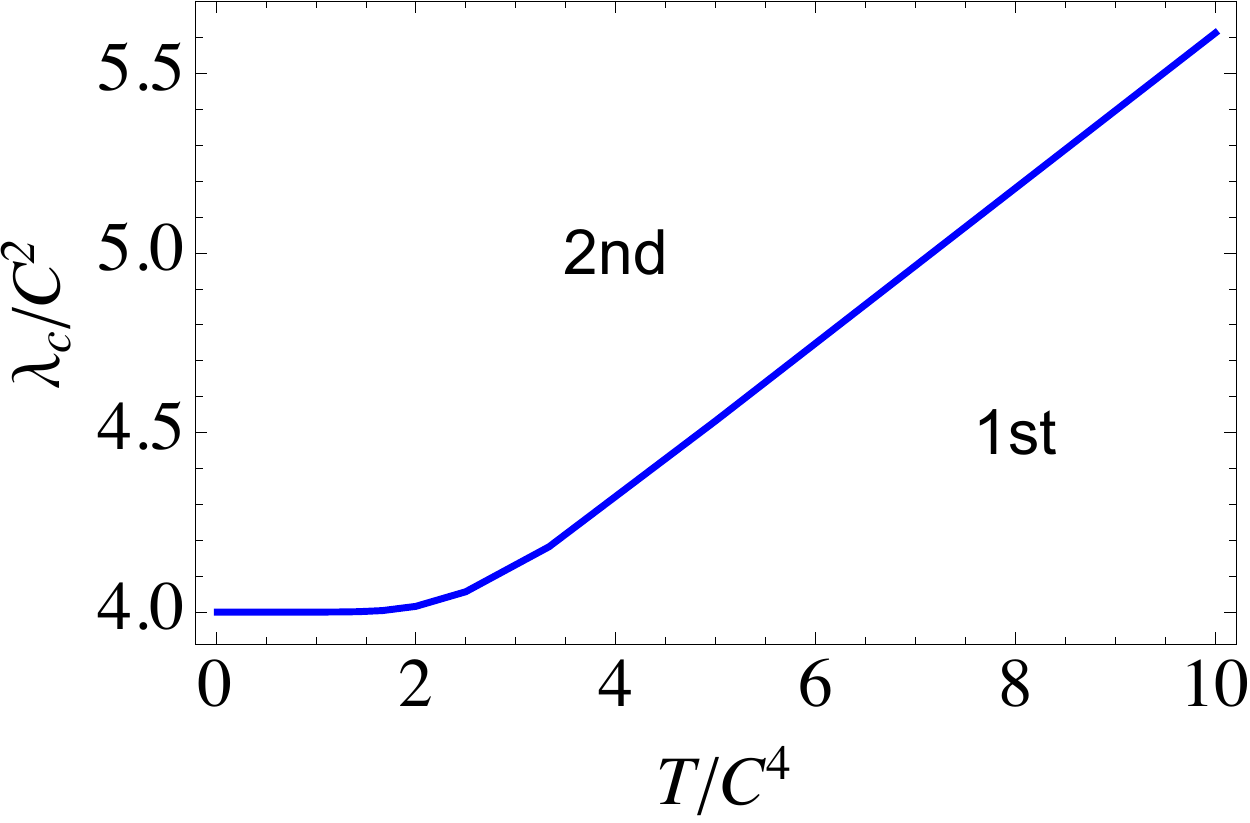}   
\caption{The critical value $\lambda_c$ at which the phase transition changes from first order to second order for $p=4$ and $q=2$. The region with $\lambda>\lambda_c$ exhibits a second order phase transition, while the region with $\lambda<\lambda_c$ exhibits a first order phase transition. Above  $T/C^4\simeq12.5$ the ferromagnetic phase disappears due to the strong thermal fluctuations.}
\label{critical lambda at finite T}
\end{figure}

In order to study the phase transition in more detail, we expand the free energy around $m=0$ to quartic order:
\bea
&&F/C^4\simeq\left({\lambda\over C^2}-{2\lambda^2\tanh(\beta\Gamma)\over C\Gamma}\right)m^2 \cr
&&+\left(
3-{2C\beta\lambda^4\over\Gamma^2}-{8C\lambda \tanh(\beta\Gamma)\over \Gamma}
+{2C\lambda^4\Gamma \tanh(\beta \Gamma)\over \Gamma^4 } \right.   \cr
&&\left.
+{2C\beta\lambda \tanh^2(\beta\Gamma)\over \Gamma^2}
\right)m^4\ ,
\eea
where we omitted the constant term $F(0)/C^4$ and higher order terms in $m$.
Let us assume that the phase transition is dominated by the quadratic and quartic terms in the expansion.
The point $m=0$ becomes perturbatively unstable when the coefficient of $m^2$ is negative, and the critical point is at $\Gamma_{c_2}$ where the coefficient of the $m^2$ term vanishes. In the following we use the subindices $c_1$ and $c_2$ to represent critical points potentially associated with first and second order phase transitions.
If the coefficient of $m^4$ is positive at $\Gamma=\Gamma_{c_2}$, $m=0$
is a global minimum for $\Gamma\ge\Gamma_{c_2}$. Therefore, the phase transition will be of second order.
If the coefficient of $m^4$ is negative, then the global minimum must be at $m=m_{c}\neq0$;
$F(m=0)>F(m_{c})$ at $\Gamma=\Gamma_{c_2}$. This means that there must be a first order phase transition at $\Gamma_{c_1}>\Gamma_{c_2}$ where 
$F(m=0)=F(m_{c}\neq0)$.
In Fig.~\ref{critical lambda at finite T}, we show the critical value $\lambda_c$, as a function of temperature, at which the order of the phase transition 
changes from first to second: a second order phase transition for $\lambda>\lambda_c$ and a first order phase transition for $\lambda<\lambda_c$. 

To understand the impact of an increasing effective penalty strength $\lambda$, consider Fig.~\ref{critical lambda at finite T} at a fixed temperature $T$. Starting from small effective penalty in the first order region, increasing $\lambda$ has the effect of eventually crossing the critical line into the second order phase transition region. From the point of view of successful QA, a second order phase transition is preferred as --- unlike the first order case --- it does not require tunneling through the free energy barrier, an event that is exponentially suppressed in the barrier width and height. At $T=0$ a first (second) order quantum phase transition is commonly associated with gap that closes exponentially (polynomially) in system size; we corroborate this in Sec.~\ref{sec:T=0 p=4} below.

Fig.~\ref{critical lambda at finite T} also shows that the higher the temperature, the larger $\lambda$ needs to be in order to affect the transition from the first order to the second order regime. Since the temperature is rescaled by $C^4$, increasing the nesting level makes the effective temperature smaller, which has the same beneficial effect for QA as increasing the effective penalty.

\subsection{Penalty strength and first-order phase transitions at zero temperature for $q=2$}

In this subsection we consider the $T=0$ case, when all the phase transitions are quantum. We study in more detail when first order phase transitions disappear for a sufficiently strong effective penalty. We only consider the $q=2$ case. 

The leading order terms of the Taylor expansions of the free energy~\eqref{free energy general pq} with $m=m_i$ around the origin $m=0$ are then:
\bes
\label{eq:Taylor}
\begin{align}
&p=3: \label{Taylor p=3}\\
&{F\over C^{3}} \simeq  {\lambda\over C} \left(1-{2C\lambda\over \Gamma}\right)m^2+2J\left(1-{3C\lambda\over \Gamma}\right)m^3\notag \\
&p=4:\label{Taylor p=4}\\
&{F\over C^{4}}\simeq  {\lambda\over C^2} \left(1-{2C\lambda\over \Gamma}\right)m^2+\left(3J-{8JC\lambda\over \Gamma}+{2C\lambda^4\over \Gamma^3}\right)m^4 \notag \\
&p\ge 5: \label{Taylor p=5} \\
&{F\over C^{p}}\simeq  {\lambda\over C^{p-2}} \left(1-{2C\lambda\over \Gamma}\right)m^2+\left({2\lambda^4\over \Gamma^3}\right)m^4 \notag
\end{align}
\ees
We omitted the $m$-independent term $F(0)/C^p$ as well as higher order terms in $m$.
As in Sec.~\ref{First-order phase transition finite T}, $m=0$ becomes perturbatively unstable when
the coefficient of the quadratic term ($m^2$) becomes negative. The critical point is
$\Gamma_{c2}=2C\lambda$. 
For $\Gamma>\Gamma_{c_2}$, the coefficient is positive and $m=0$ is perturbatively stable. For $\Gamma<\Gamma_{c_2}$, $m=0$ is perturbatively unstable. 

It is important to note that this quadratic term does not 
exist if the effective penalty vanishes ($\lambda=0$); in this case the phase transition is always of first order. It is the quadratic penalty terms that enable the second order phase transitions. Whether the first order phase transitions persist or not depends on the higher order terms.
If the coefficient of the next lowest order term beyond $m^2$ is positive
at $\Gamma=\Gamma_{c_2}$, then the phase transition will be of second order,
and if it is negative then the phase transition will be of first order.
Of course, this estimation of the first order phase transition is based on
the assumption that the lowest and the second lowest terms in the Taylor expansion determine the phase transition. 
As we will show in this section, this assumption is in fact not valid for higher $p$.

\subsubsection{$p=3$}

\begin{figure}[t]
\includegraphics[width=0.45\textwidth]{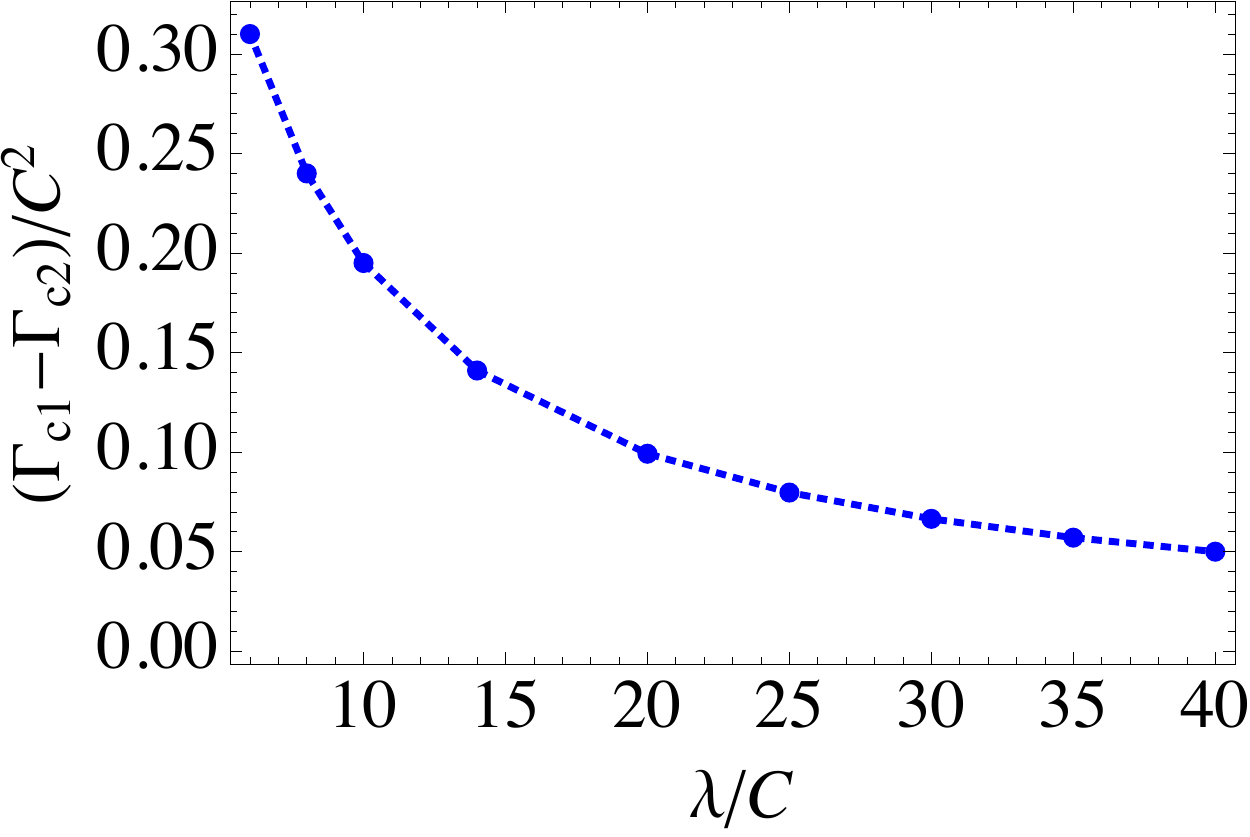}
\caption{The $p=3$ case.
The difference of the critical values  of the transverse field $(\Gamma_{c_1}-\Gamma_{c_2})/C^2$ as a function of $\lambda/C$. Since $\Gamma_{c_1}>\Gamma_{c_2}$,
there is only a first order phase transition.
}
\label{p3gamGam}
\end{figure}

Let us first consider $p=3$. From Eq.~\eqref{Taylor p=3}, both of the coefficients of $m^2$ and $m^3$ are positive for $\Gamma\ge 3C\lambda$, so that $m=0$ is the global minimum.
For $\Gamma_{c_2}=2C\lambda<\Gamma< 3C\lambda$, the coefficient of $m^3$ becomes negative while that of $m^2$ is still positive. This suggests that there exists a critical 
value $\Gamma_{c_2}<\Gamma_{c_1}$ and for $\Gamma\le \Gamma_{c_1}< 3C\lambda$, the global minimum shifts from $m=0$ to $m=m_c\neq0$. The critical value $\Gamma_{c_1}$
is determined by $F(0)=F(m_c)$. This is a first order phase transition since $m=0$ and $m=m_c$ are separated by a potential barrier.
Since $\Gamma_{c_2}<\Gamma_{c_1}$ independent of the value of $\lambda$, the first order phase transition is inevitable. We show the 
difference between $\Gamma_{c_1}$ and $\Gamma_{c_2}$ in Fig.~\ref{p3gamGam}.

\subsubsection{$p=4$}
\label{sec:T=0 p=4}

For $p=4$, the coefficient of the subleading term, namely the coefficient of $m^4$ in Eq.~\eqref{Taylor p=4}
at the critical point $\Gamma_{c_2}=2C\lambda$ is $\lambda/C^2-4$. The coefficient is negative for ${\lambda/ C^2}<4$
which means that there is only a first order phase transition at $\Gamma_{c_2}<\Gamma_{c_1}$. It is positive for
${\lambda/ C^2}>4$ and a second order phase transition takes place to destabilize $m=0$.
While this conclusion is based on a Taylor expansion analysis, a full numerical analysis shows that there is no first order phase transition for $p=4$ above ${\lambda/ C^2}=4$. Indeed, this is already clear from Fig.~\ref{fig:p=4}, which is for $T/C^4=0.01$; the only change at $T=0$ is that the vertical axis scale is somewhat compressed.

\begin{figure}[t]
\includegraphics[width=0.45\textwidth]{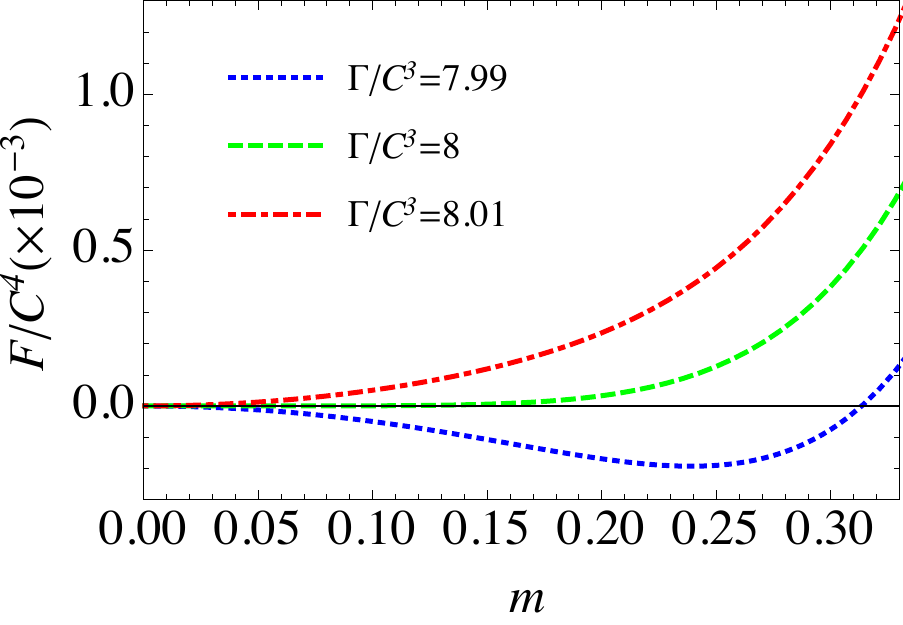}
\caption{ 
Free energy around the critical point for $\lambda/C^2=4$, $\Gamma/C^3=8$ for $p=4$.}
\label{p4crit}
\end{figure}

Figure~\ref{p4crit} shows the free energy as a function of $m$ near the critical point $\Gamma/C^3=8$ for $p=4$ and $\lambda/C^2=4$ where the first-order phase transition disappears.
One can see that the phase transition is of second order even though the interaction has a higher order power ($p>2$). The point at which $\lambda/C^2=4$ and $\Gamma/C^3=8$ is where the coefficients of the first two leading terms $m^2$ and $m^4$ vanish simultaneously, and the next order leading term ($m^6$) has a positive coefficient. Above this value of $\lambda$ the phase transition is always of second order, and below it is of first order.

\begin{figure}[t]
  \includegraphics[width=0.45\textwidth]{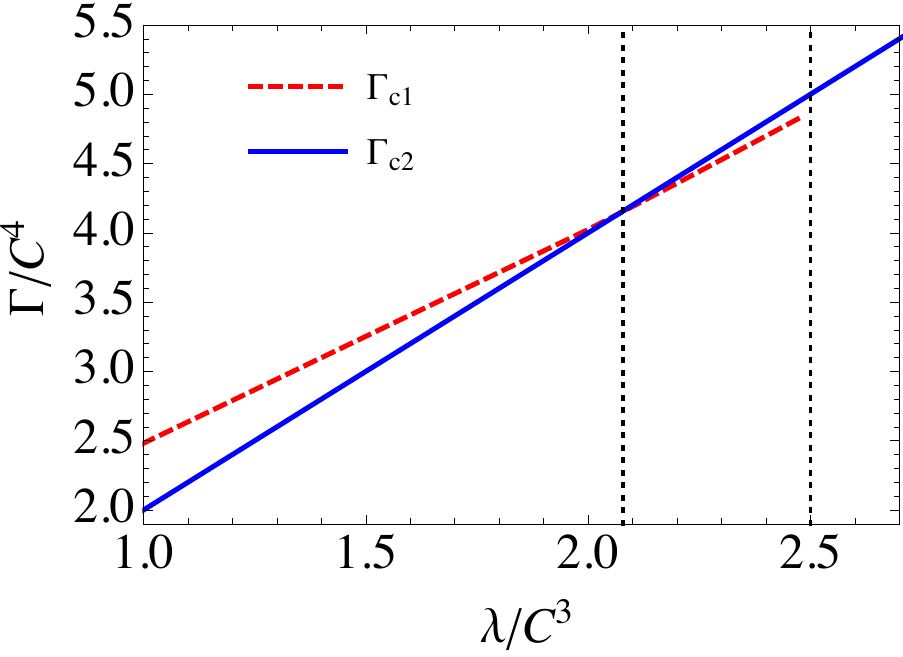}
\caption{
The critical values $\Gamma_{c_1}/C^4$ and $\Gamma_{c_2}/C^4$ as a function of $\lambda/C^3$ for $p=5$. For $\lambda/C^4<2.078$, $\Gamma_{c_1}>\Gamma_{c_2}$. Therefore, there is a first order phase transition. For $2.078<\lambda/C^4<2.5$ the first order phase transition and the second order phase transition coexist. 
For $2.5<\lambda/C^4$, there is only a second order phase transition.
}
\label{plp5lGmFfirst}
\end{figure}

\subsubsection{$p\geq 5$}

The Taylor expansion for $p=5$, Eq.~\eqref{Taylor p=5}, suggests that the subleading $m^4$ term is always positive.
This does not mean that the phase transition is always of second order.
In fact, since there is a single first order phase transition for $p\ge 3$ in the absence of the effective penalty, by continuity there must be a first order phase transition when $\lambda$ is very small. This suggests that higher order terms in $m$ become relevant for these first order phase transitions.
In Fig.~\ref{plp5lGmFfirst}, we plot $\Gamma_{c_1}$ and $\Gamma_{c_2}$ as a function of the effective penalty coupling $\lambda/C^3$. They cross at $\lambda/C^3 \approx 2.078$.
When $\lambda/C^3< 2.078$, there is a single first order phase transition between $m=0$ and $m=m_c>0$.
For $ 2.078<\lambda/C^3$, $m=0$ becomes perturbatively unstable and there is a second order phase transition from $m=0$ to 
$m=m_c>0$. There are two different cases after the second order phase transition. 
For $ 2.078<\lambda/C^3<2.5$, there is a first order phase transition at $\Gamma_{c_1}<\Gamma_{c_2}$ and the order parameter changes discontinuously from
$m_{c}$ to {$m>m_{c}$}. On the other hand, for $2.5<\lambda/C^3$
there is no phase transition other than the second order phase transition from $m=0$ to $m=m_{c}$.

\begin{figure}[t]
   \includegraphics[width=0.45\textwidth]{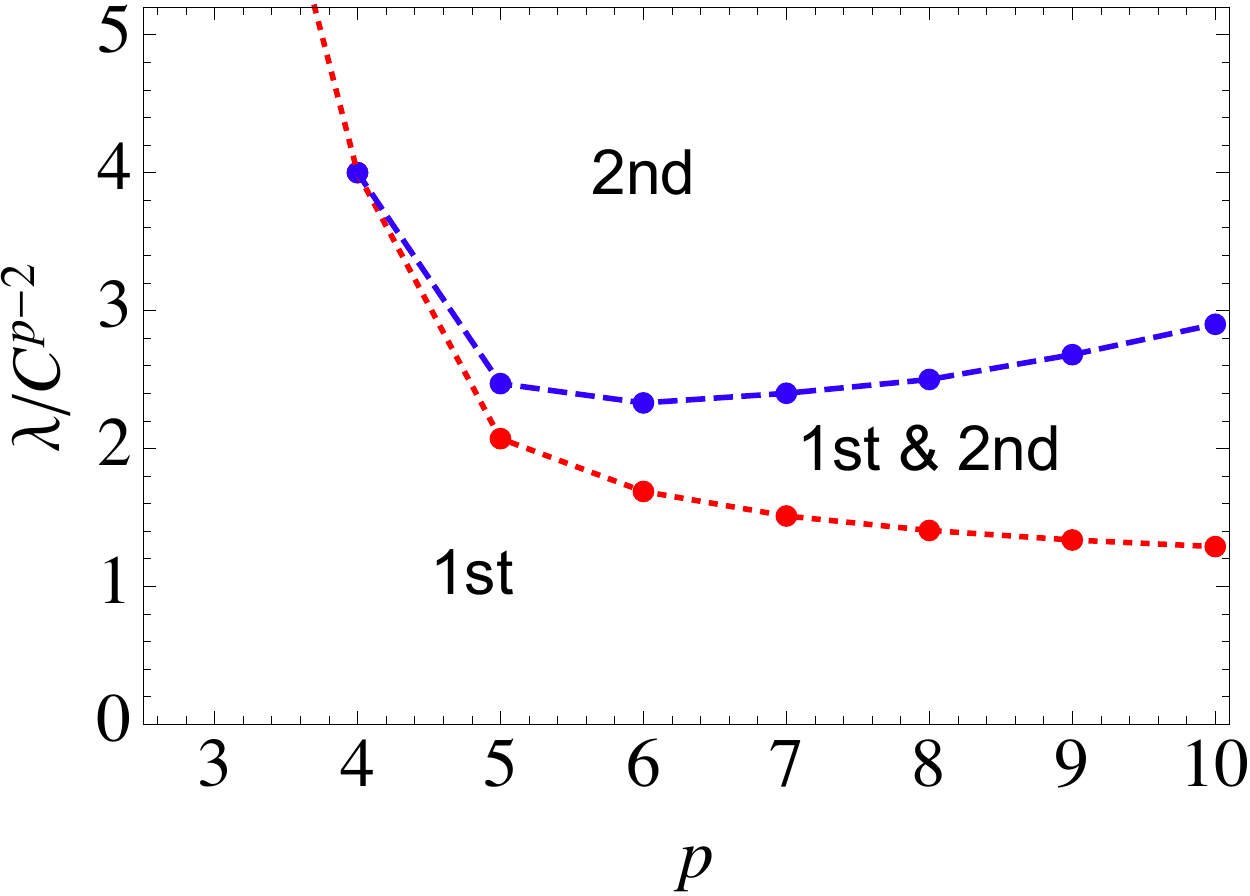} 
\caption{
The order of the phase transitions for various values of $p$ and $q=2$.
For $p=3$ there is a unique first order phase transition. For $p=4$, there is a critical value of $\lambda$ above which the phase transition becomes of second order.
For $p\ge5$, there is an intermediate region where the first and second phase transitions coexist. 
For $p\ge4$, all the phase transitions turn into second order phase transitions for sufficiently large $\lambda$.
}
\label{phasegeneral}
\end{figure}

This behavior is general for $p\ge5$. In Fig.~\ref{phasegeneral}, we show the order of the phase transitions for various values of $p$ and $\lambda$. As we saw earlier, only the first order phase transition exists for $p=3$. For $p=4$, the first order phase transition changes into a second order phase transition at $\lambda/C^{p-2}=4$. For $p\ge5$ there is a parameter region where the first order and the second order phase transitions coexist. The first order phase transition disappears above a certain $p$-dependent value of $\lambda/C^{p-2}$.

\subsection{Energy gap from instanton analysis}

While the width and height of the free energy barrier are useful measures for a qualitative inference of the change of the energy gap~\cite{Matsuura:2016aa}, one can estimate the energy gap more accurately in terms of the eigenstates of the effective Hamiltonian.
The energy gap is given by the instanton solution in the Euclidean path integral which connects two different minima of the free energy.
When a transition between two minima happens instantaneously, the energy gap can be approximated by the overlap of two eigenstates of the effective Hamiltonian at each local minimum~\cite{Jorg:2010qa,MNAL:15,Matsuura:2016aa}.

\begin{figure}[t]
 \includegraphics[width=0.45\textwidth]{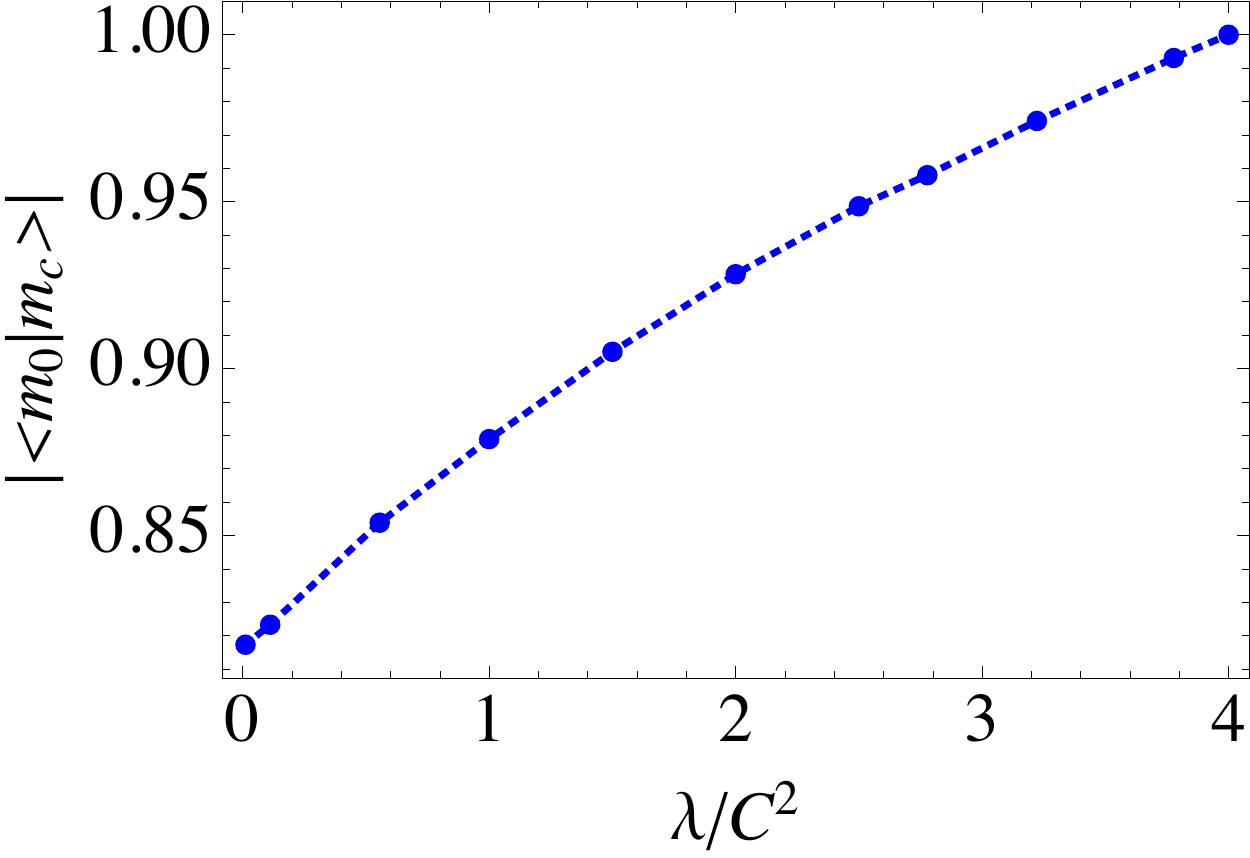}   
\caption{The overlap of the $m=m_0=0$ and $m=m_{\rm c}$ eigenstates of the effective $T=0$ Hamiltonian~\eqref{eq:effHam p=4 q=2}, as a function of the effective penalty $\lambda$ for $p=4, q=2$. This overlap determines the closing of the gap, as per Eq.~\eqref{eq:Delta}.}
\label{pgvsM}
\end{figure}

Let us discuss this viewpoint in some detail. At the critical point, there are two minima characterized by the order parameters $m=m_0=0$ and $m=m_{c}>0$.
A first-order phase transition happens between $m_{0}$ and $m_{c}$.
The last term of the free energy
~\eqref{general free energy eq}
, which represents quantum effects as it involves $\Gamma$, 
takes the form of $\cosh(\beta \epsilon)$
with 
$\epsilon=\sqrt{(pJ C^{p-1} m^{p-1}+q\lambda C^{q-1} m^{q-1})^2+\Gamma^2}$ after using $m=m_i$. 
This contribution to the energy can be reproduced  by the linearized
Hamiltonian 
which is obtained by replacing every power $>1$ of the Pauli matrices $\sigma^{z}_{ic}$ by the expectation value $m$:
\bea
H_{\rm eff}(m)=-(pJ C^{p-1} m^{p-1}+q\lambda C^{q-1} m^{q-1}) \sigma^{z}-\Gamma  \sigma^{x}\ .
\label{eq:effHam p=4 q=2}
\eea
Let $| m_{0} \rangle$ and $| m_{c} \rangle$ be the ground states of
$H_{\rm eff}(m_{0})$ and $H_{\rm eff}(m_{c})$, respectively.
Then the energy gap $\Delta$ is (for a detailed justification see Appendix B of Ref.~\cite{Matsuura:2016aa}):
\bea
\Delta\sim |\langle m_{0} | m_{c} \rangle|^{NC}
\label{eq:Delta}
\eea
Figure~\ref{pgvsM} shows the wavefunction overlap as a function of the effective penalty $\lambda$. It reaches $1$ at the critical penalty strength
$\lambda/C^2=4$, which implies that the gap ceases to decrease exponentially as a function of the system size. This agrees with the conclusions we drew above about the phase transition changing from first order (corresponding to an exponentially decreasing gap) for $\lambda/C^2< 4$ to second order for $\lambda/C^2\geq4$.

\subsection{$C$ dependence and $\lambda$ dependence}
\label{sec:C-dep}

As can be seen from Eq.~\eqref{general free energy eq},
the free energy scales as  $C^{\text{max}(p,q)}$ for large $C$. 
Since the occupancy rate of the excited states is suppressed by the factor
$e^{-\beta N F(m)}$, thermal fluctuations are efficiently suppressed in terms of the nesting level $C$.
This is an essential feature required for scalable error correction, and NQAC satisfies this condition.
Note that the free energy in PQAC \cite{PAL:13,PAL:14,Mishra:2015} scales linearly in $C$ \cite{MNAL:15,Matsuura:2016aa}. Therefore, NQAC removes thermal noise more efficiently than PQAC.

While increasing $C$ is beneficial for suppressing thermal fluctuations, it may enhance Landau-Zener transitions at the critical point in a first order phase transition:
the energy gap is suppressed by a power of $C$ [Eq.~\eqref{eq:Delta}]. 
On the other hand, the wavefunction overlap $|\langle m_{0} | m_{c} \rangle|$ in Eq.~\eqref{eq:Delta} can be increased by increasing the penalty coupling $\lambda$, as shown Fig.~\ref{pgvsM}.
Therefore, one can adjust the parameters $\{C,\lambda\}$ so that both thermal fluctuations and Landau-Zener transitions are efficiently suppressed.


\section{Antiferromagnetic case}
\label{antiferromagnetic case}

In this section we study fully antiferromagnetic systems, where the coupling satisfies $J<0$. Unlike the ferromagnetic case, the long-range antiferromagnetic system is frustrated and thus exhibits different behavior.

The free energy and the saddle point equations are the same as those in the ferromagnetic case, i.e., Eq.~\eqref{general free energy eq}, the only difference being that $J$ is replaced by $-J (<0)$. The penalty $\gamma$ coupling (and hence also the effective penalty $\lambda$) between physical qubits within a logical qubit remains ferromagnetic as it is designed to introduce resilience against bit flips in a logical qubit. It tends to induce a ferromagnetically ordered state for each logical qubit, i.e., $m_i\neq0$. 

Let us assume that the total number of logical qubits is even and that the effective penalty strength is large, i.e., $\gamma \gg J$, so that the logical qubits are ordered 
In this case, 
there are two possible phases in the model with antiferromagnetic interactions.
The first is the paramagnetic phase, where each logical qubit has a vanishing expectation value $m_i=0$ and consequently the total magnetization also vanishes: $m=0$.
This phase is realized when the temperature and/or the transverse field are large.
The second phase is where each logical qubit is ordered, i.e., $m_i=\pm n$, while the total magnetization again vanishes: $m=0$.
We call this a ``locally ordered phase"; 
it appears when the effect of $\lambda$  is stronger than that of the temperature or the transverse field.
In the following we analyze these phases at or near the ground state, analytically as well as numerically. In addition, there can be metastable states as in the ferromagnetic example, which we analyze in the appendix.

Since both of the phases take $m=0$, the value of $p$ does not determine the order of the phase transition, which is different from the ferromagnetic case. Instead, the order of the phase transition is controlled by $q$. All the $J$ dependence disappears from the mean field free energy [Eq.~\eqref{general free energy eq}] and the $p$-dependence appears only in the scaling of parameters in $C$ discussed in Sec.~\ref{sec Problem_formulation}.

\begin{figure}[t]
 \includegraphics[width=0.45\textwidth]{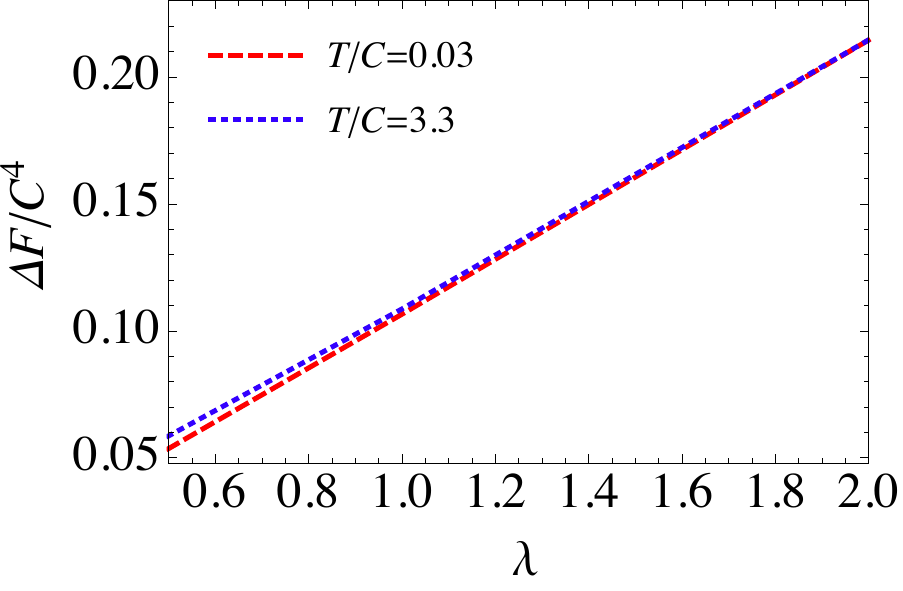}
\caption{Free energy barrier in the antiferromagnetic case between two phases at a first-order phase transition, at $T/C=0.03$ and $3.3$, for $p=q=4$. Contrast with Fig.~\ref{fig:p=4} for the ferromagnetic case.
}
\label{gvsDFq4b01p}
\end{figure}

Assuming that the total number of spins $N$ is even, the free energy~\eqref{general free energy eq} reduces to:
\bea
\label{general 000 free energy AF}
F&=&{(q-1)\over 2}\lambda C^{q} (n^{q}+(-n)^{q})
+\\
&-&{C\over \beta }
 \log\left. \bigg( 2\cosh \beta   \sqrt{(q\lambda C^{q-1} n^{q-1})^2+\Gamma^2}\right) \notag
\eea
Let us first consider the $q=2$ case. The expression for the free energy is the same as for the ferromagnetic case with two-body interactions $p=2$ [Eq.~\eqref{free energy eq}], with the local (logical qubit) order parameter $n={1\over C } \langle \sum_{c}\sigma^{z}_{ic} \rangle$ replacing the global order parameter $m={1\over NC }\langle  \sum_{i,c}\sigma^{z}_{ic}\rangle$ appearing in the ferromagnetic case. I.e., the free energy of the ferromagnetic system with the order parameter $m$ is the same as that of the antiferromagnetic system with the different order parameter $n$.
Therefore, the latter has a second-order phase transition.
It should be noted that the conventional $p$-body antiferromagnetic Ising model $(C=1)$ does not have a phase transition for $p$ even.\footnote{The classical energy of antiferromagnetic interactions $+JN(\sum_{i=1}^N \sigma_i^z /N)^p~(J>0)$ becomes smallest when $\sum_{i=1}^N \sigma_i^z$ is 0, the paramagnetic state.  This state has the largest entropy since the number of possible spin configurations is largest. Thus the free energy is lowest for the paramagnetic state irrespective of the temperature for the classical model, i.e., no phase transition. The situation is essentially the same in the quantum case with a transverse field. For $p$ odd, the antiferromagnetic model becomes equivalent to a ferromagnetic model under the spin inversion $\{\sigma_i^x, \sigma_i^y, \sigma_i^z\} \mapsto \{\sigma_i^x, -\sigma_i^z, -\sigma_i^y\}$.
} 

The phase transition point is determined in the same way as in the ferromagnetic case, where we derived Eq.~\eqref{ferrosaddle}. The critical parameter values satisfy $2\lambda C\tanh(\beta \Gamma)/\Gamma=1$, which is $p$-independent. The paramagnetic and locally ordered phases exist in the respective regions
\bes
\begin{align}
&2\lambda C\tanh(\beta \Gamma)/\Gamma<1~~~\ :  \quad m_i = 0 
\\
&2\lambda C\tanh(\beta \Gamma)/\Gamma>1~~~\ :  \quad m_i \equiv \pm n \neq 0 
\label{locally broken condition}
\end{align}
\ees
 It is clear from these equations that high temperature and large transverse field favor the paramagnetic phase (all $m_i=0$), while large local ferromagnetic coupling $\lambda$ favors the locally ordered phase (all $m_i\neq0$).
In addition, we see that a larger value of $C$ also favors the locally ordered phase. This is because 
$C$ aligned spins behave as a single spin-$C$ vector, thus increasing the effective ferromagnetic coupling by $C^2$.

It is straightforward to extend the analysis to a general $q$.
For even $q (\ge 4)$, the phase transition is of first order,
and for odd $q (\ge 3)$ there is no phase transition since the state with the lowest value of the free energy always has $m_i\neq0$.

Figure~\ref{gvsDFq4b01p} shows the height of the free energy barrier at the transition point for $q=4$.
Clearly, a larger value of the penalty coefficient $\lambda$ has a higher barrier height, in sharp contrast to the ferromagnetic case (Fig.~\ref{fig:p=4}).
The $T/C$-dependence is very weak.
The jump in the order parameter $\Delta m$ at the transition does not depend on $\lambda$ and $T/C$ within the numerical precision of our simulations.

\section{Hybrid NQAC-PQAC strategy}
\label{Penalty term in the nested QAC}

The penalty term in NQAC prevents transitions from the code space by imposing ferromagnetic couplings between physical qubits within a logical qubit. This is not the only way to impose a penalty. 
For instance, one can introduce an independent and designated penalty qubit for each logical qubit, and connect this penalty qubit and the physical qubits in the logical qubit ferromagnetically. We refer to this method as penalty quantum annealing correction; it was proposed in Ref.~\cite{PAL:13} and further studied in Refs.~\cite{PAL:14,Mishra:2015}. As shown in Refs.~\cite{MNAL:15,Matsuura:2016aa}, phase transitions can disappear or be significantly weakened when the penalty coupling is sufficiently large.

In this section we consider a hybrid penalty-nested QAC model, and introduce designated penalty qubits into NQAC. 

\subsection{Ferromagnetic case}

Let us explicitly write the Hamiltonian for the case of the $p$-spin ferromagnetic model,
\bea
H
&=&-{NJ} \left({1\over N}\sum_{i=1}^{N}\sum_{c_i=1}^{C}\sigma_{ic_i}^{z}\right)^{p}
-\lambda\sum_{i=1}^{N}\left(\sum_{c_i=1}^{C} \sigma_{ic_i}^{z}\right)^{q} \cr
&&-\Gamma \sum_{i=1}^{N}\sum_{c_i=1}^{C}\sigma_{ic_i}^{x}
-\eta\sum_{i=1}^N\left(\sum_{c_i=1}^C\sigma_{ic_i}^{z}\right)\sigma^{z}_{i0}\ ,
\label{ferro-hamil-penal}
\eea
where each designated penalty qubit (labeled $i0$) couples ferromagnetically via $\sigma^{z}_{i0}$ to the physical qubits within a logical qubit labeled by $i$ with a coupling strength $\eta$.
The corresponding free energy is
\bea 
\label{eq:F-hybrid}
F&=&
(p-1)JC^{p}m^p+(q-1)\lambda {C^q\over N}\sum_{i} m^q_i  \cr
&&-{1\over \beta N}
\log\left(
\sum_{a_i=\pm}
\left[\prod_i 2\cosh  \beta v^{a_i}_{i} \right]^C  
\right)
\eea
where
\begin{align}
&v^{\pm}_{i}=C^{p-1}\times\\
&\sqrt{[p J m^{p-1} +q\lambda C^{q-p} m^{q-1}_i \pm (\eta/C^{p-1})]^2+(\Gamma/C^{p-1})^2}\ . \notag
\end{align}
The penalty coupling is rescaled as
$\eta\mapsto {\eta/ C^{p-1}}$, and the effective temperature decreases as $T/C^{p}$ (from the partition function $Z=e^{-N\beta F}$), which serves to suppress thermal excitations.

Let us now focus our attention on the ground state at zero temperature. Since the system is ordered in the ground state, the local magnetization of each logical qubit assumes the same value as that of the total system, i.e., $m=m_i$.
Then Eq.~\eqref{eq:F-hybrid} yields:
\begin{align}
&F=(p-1)JC^{p}m^{p}+(q-1)\lambda C^{q}m^{q}-C^{p}\times\\
&\sqrt{(pJm^{p-1}+q\lambda C^{q-p}m^{q-1}+(\eta/ C^{p-1}))^2+(\Gamma/C^{p-1})^2}\ . \notag 
\end{align}
To study the stability of the paramagnetic state $m=0$, we perform a Taylor expansion of $F$ around $m=0$ [compare to Eq.~\eqref{eq:Taylor} for the NQAC case].
For $p>q$:
\bea
F/C^p\simeq-{q(\eta\lambda)C^{q-p}\over \sqrt{\eta^2+\Gamma^2}}m^{q-1} +\mathcal{O}(m^{q})
\eea
where we neglected a constant term $F(0)$.
For $p=q$:
\bea
F/C^p\simeq-{q(\eta(J+\lambda))\over \sqrt{\eta^2+\Gamma^2}}m^{q-1}  +\mathcal{O}(m^{q})
\eea
If there is no nesting penalty coupling (i.e., $\lambda=0$) then:
\bea
F/C^p\simeq-{p(J\eta)\over \sqrt{\eta^2+\Gamma^2}}m^{p-1}  +\mathcal{O}(m^{p})
\eea
In all cases, we see that the stable state is ferromagnetic, $m\ne 0$.

\begin{figure}[t]
 \includegraphics[width=0.45\textwidth]{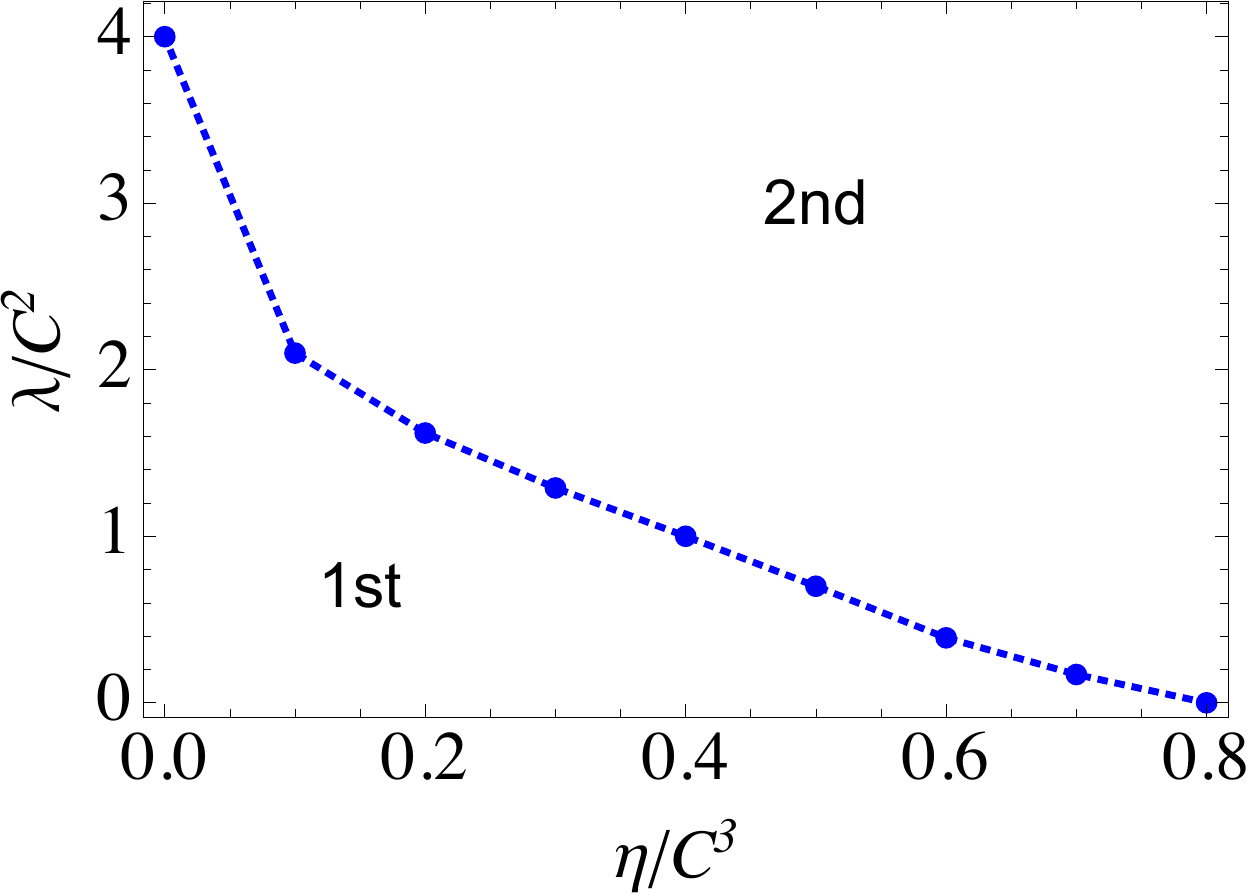}
\caption{Critical values of the penalty strength $\eta/C^3$ and $\lambda/C^2$ for
 $T=0,p=4,q=2, J=1$, for the hybrid NQAC-PQAC strategy and ferromagnetic coupling. The phase transition is of first order
 below the blue dotted line and of second order above it. Since the second order phase transition region is preferred, increasing $\eta$ in the hybrid NQAC-PQAC strategy allows for a lower value of $\lambda$ than in the pure NQAC case.}
\label{etalamCritp4}
\end{figure}

Figure~\ref{etalamCritp4} shows the critical line for $T=0,p=4$ and $q=2$.
In the parameter region ($\eta,\lambda$) below the line the system undergoes a first order phase transition, while above the line it undergoes a second order phase transition. The critical value of $\lambda$ decreases rapidly when $\eta$ is turned on and then decreases almost linearly above $\eta/C^3\sim 0.1$. 
As expected, the critical value of $\eta/C^3$ at $\lambda/C^2=0$ reproduces the result for penalty quantum annealing correction found in Ref.~\cite{MNAL:15}. The main point is that the two types of penalties are complementary, in the sense that increasing one (say $\eta$) allows for a smaller value of the other (say $\lambda$). 

\subsection{Antiferromagnetic case}

Assuming $\gamma \gg J$ as in Sec.~\ref{antiferromagnetic case} to ensure ordering of the logical qubits, the ground state of the antiferromagnetic system has $m_i=\pm n$, $m=0$. 
The free energy is 
\bea
F&=&
(q-1) \lambda C^{q}  n^{q} - {1\over \beta N} \cr
&&
\log\left(\left[ 2\cosh  \beta\sqrt{(q\lambda C^{q-1} n^{q-1} +\eta)^2+\Gamma^2}\right]^{C} \right. \cr
&&
\left. +\left[ 2\cosh  \beta\sqrt{(q\lambda C^{q-1} n^{q-1} -\eta)^2+\Gamma^2}\right]^{C}   \right)^{N}\ .
\eea
At zero temperature this becomes:
\bea
\label{eq:41}
F&=&
(q-1) \lambda C^{q}  n^{q} -  \cr
&&~- \sqrt{(q\lambda C^{q} |n|^{q-1} +C\eta)^2+(C\Gamma)^2} 
\eea
The free energy in this case is identical to the free energy of the ferromagnetic system with a designated penalty term discussed in Refs.~\cite{MNAL:15,Matsuura:2016aa}, after the rescaling $\eta \mapsto C\eta$, $\Gamma \mapsto C\Gamma$.%
\footnote{Compare Eq.~\eqref{eq:41} to Eq.~(3b) in Ref.~\cite{MNAL:15} and in addition identify $K\mapsto C$, $m_k \mapsto n$, $q\mapsto p$, $\gamma \mapsto \eta$, and set $\lambda = C^{1-q}$.}
This system has a rather remarkable feature.
When $\eta=0$, there is, respectively, a second and a first order phase transition for $q=2$ and $q\ge3$.
At finite $\eta$, the second order phase transition for $q=2$ disappears at $T=0$.
For $q\ge3$ and $T=0$, there is a first order phase transition for $\eta$ below a critical value $\eta_c$. However, the strength of the first order phase transition 
becomes weaker as $\eta$ increases and above the critical value $\eta_c$, the first order phase transition disappears.
At finite temperature, the first order phase transition exists at any value of $\eta$. However, the strength of the phase transition significantly weakens as $\eta$ increases.
More details can be found in Refs.~\cite{MNAL:15,Matsuura:2016aa}.

To summarize, the penalty coupling $\lambda$ and the nesting level $C$ reduce or effectively remove the thermal fluctuations. 
While the energy gap closing associated with increased nesting level $C$ (recall Sec.~\ref{sec:C-dep})
still exists in NQAC, 
it can be removed or significantly weakened by introducing designated penalty qubits and their coupling $\eta$, as illustrated in Fig.~\ref{etalamCritp4}.

\section{Discussion and Conclusions}
\label{sec:discussion}

We have studied nested quantum annealing correction (NQAC)~\cite{vinci2015nested,Vinci:2017ab} and the combination of NQAC and penalty quantum annealing correction (PQAC)~\cite{PAL:13,PAL:14} at zero and finite temperatures.
While PQAC (studied theoretically in Refs.~\cite{MNAL:15,Matsuura:2016aa}) has designated penalty qubits which serve to induce a symmetry breaking field,
NQAC does not have designated penalty qubits and therefore
the penalty coupling itself does not induce symmetry breaking.
Nevertheless, we have shown that first-order phase transitions can be removed or significantly weakened when the interaction strength between physical qubits within a logical qubit is chosen appropriately.
Moreover, the nested structure works to efficiently reduce thermal fluctuations: for $p$-body coupling, the temperature $T$ is reduced effectively to $T/C^p$.
Therefore, NQAC helps to stabilize the ground state by reducing thermal fluctuations, and serves as a means to address the `temperature scaling law' problem identified in Ref.~\cite{Albash:2017ab}, that in order to serve as optimizers, quantum annealer temperatures must be appropriately scaled down with problem size.

While a larger value of the penalty coefficient is beneficial for keeping the state in the ground state, it may cause a different problem. A larger value of the penalty term creates local minima in the free energy for excited states, as detailed in the Appendix.
Once the system state becomes excited and trapped by one of those metastable states, it may be difficult to escape to the ground state if the penalty coupling is too strong.
This is a difficult problem to analyze, and we leave the evaluation of transitions between metastable states and the ground state for future investigations.

The success of NQAC sheds some light on the computational efficiency and the local structure of the problem.
In PQAC, the order parameters are the expectation values of total spins~\cite{MNAL:15,Matsuura:2016aa}.
On the other hand, NQAC has an extensive number of order parameters ($m_i$) in addition to the expectation value of the total spin ($m$).
When the penalty coupling is strong, the behavior of the free energy is determined by the local order parameters $m_i$.
Therefore, by tuning the \emph{local} physics (the strength of the penalty coupling $\gamma$), one can change the 
structure of the phase transition of the \emph{whole} system.
In our examples, even when the original problem has a first-order phase transition, e.g., for the case of $p\ge3$, by encoding qubits properly in NQAC with $q=2$, one can change the order of the phase transition into second.
It is an interesting problem how these results generalize beyond simple mean-field models with uniform interactions.

\section{acknowledgement}

 The work of WV and DL was (partially) supported under ARO grant number W911NF-12-1-0523, ARO MURI Grant Nos. W911NF-11-1-0268 and W911NF-15-1-0582, and NSF grant number INSPIRE-1551064. This research is based upon work partially supported by the Office of the Director of National Intelligence (ODNI), Intelligence Advanced
Research Projects Activity (IARPA), via the U.S. Army Research Office
contract W911NF-17-C-0050. The views and conclusions contained herein are
those of the authors and should not be interpreted as necessarily
representing the official policies or endorsements, either expressed or
implied, of the ODNI, IARPA, or the U.S. Government. The U.S. Government
is authorized to reproduce and distribute reprints for Governmental
purposes notwithstanding any copyright annotation thereon.



\appendix

\section{Derivation}
\label{Appendix derivation}

In this Appendix, we derive the free energy of NQAC by using mean field theory.
We consider an NQAC Hamiltonian 
\bea
H=H_{X}+H_{Z}
\eea
where, as in Eq.~\eqref{eq:H-general}:
\bes
\begin{align}
\label{eq:HZ}
H_{Z}&\equiv -{JN} \left({1\over N}\sum_{i=1}^{N}\sum_{c_i=1}^{C}\sigma_{ic_i}^{z}\right)^{p}
-
\lambda\sum_{i=1}^{N}\left(\sum_{c_i=1}^{C} \sigma_{ic_i}^{z}\right)^{q} \\
\label{eq:HX}
H_{X}&\equiv
-\Gamma \sum_{i=1}^{N}\sum_{c_i=1}^{C}\sigma_{ic_i}^{x}\ .
\end{align}
\ees
From now on we replace the double subscript $ic_i$ by $ia$ to simplify the notation, and keep in mind that $a$ enumerates the physical qubits within the $i$-th logical qubit.

Skipping many steps which parallel the calculation in Appendix A of Ref.~\cite{Matsuura:2016aa}, the partition function $Z=\tr(e^{-\beta H})$ is computed via the Suzuki-Trotter decomposition:

\onecolumngrid
\bea
Z&=&\lim_{M\to\infty}\tr \left(
e^{-{\beta\over M}H_{X}}
e^{-{\beta\over M}H_{X}}
\right)^{M}\cr
&=&
\lim_{M\to\infty}\prod_{i}^{N}
 \prod_{\al}^{M} \int dm_{\al}  d\tilde{m}_{\al} dm_{i\al}  d\tilde{m}_{i\al} 
 \exp\left[i\tilde{m}_{\al}\left(N C m_{\al}-\sum_{i=1}^{N}\sum_{a=1}^{C}\sigma_{ia}^{z}(\al)\right)
+{\beta J \over M } N C^p m_{\al}^{p} \right.+ \cr
&+&\left.\sum_{i} \left(
i\tilde{m}_{i\al}\left(C m_{i\al}-\sum_{a=1}^{C} \sigma_{ia}^{z}(\al)\right)
+{\beta \lambda  \over M} C^q m_{i\al}^{q}\right)+{\beta \Gamma \over M}\sum_{i=1}^{N}\sum_{a=1}^{C}\sigma_{ia}^{x}(\al) 
\right]  \prod \langle \sigma | \sigma \rangle\,,
\eea
where $M$ is the Trotter number and
\bea
 \prod \langle \sigma | \sigma \rangle\equiv
\prod_{c=1}^{C}
\langle \sigma^{z}_{ic}(\al)|\sigma^{x}_{ic}(\al)\rangle \langle \sigma^{x}_{ic}(\al) | \sigma^{z}_{ic}(\al+1) \rangle
.
\eea
We use the static ansatz  where all the variables are Trotter index $\al$ independent; $m_{\al}=m, \tilde{m}_{\al}= \tilde{m},m_{i\al}=m_i, \tilde{m}_{i\al}= \tilde{m}_{i}$.
We also rescale $(\tilde{m},\tilde{m}_i)$ to  ${1\over M}(\tilde{m},\tilde{m}_i)$. Then, in the thermodynamic limit $N\to\infty$,
\bea
Z&=&
\exp\left(i  NC \tilde{m}m  +{\beta J N} C^{p} m^{p}+i    C\sum_{i} \tilde{m}_i m_i+C^{q} \sum_{i}{\beta\lambda } m_{i}^{q} \right) 
\left(
\Tr \exp\left(
-\sum_{i=1}^{N} i\tilde{m} \sigma_{ia}^{z}
-i\tilde{m}_i  \sigma_{ia}^{z}
+ \beta\Gamma \sigma_{ia}^{x}
\right)
\right)^C
 \cr
&=&
\exp\left(i N C \tilde{m}m  +{\beta J N } C^{p} m^{p}+i   C \sum_{i} \tilde{m}_i m_i+ C^{q}\sum_{i}{\beta \lambda } m_{i}^{q} \right) 
\left(
\prod_{i} (e^{\sqrt{(- i\tilde{m} -i\tilde{m}_i)^2+(\beta\Gamma)^2}}+
e^{-\sqrt{(- i\tilde{m} -i\tilde{m}_i)^2+(\beta\Gamma)^2}}
)
\right)^{C}
\cr
&&
\eea
The saddle point equations for $m$ and $m_i$ are
\bes
\bea
i\tilde{m}+\beta p J C^{p-1} m^{p-1} =0\ ,\\
i\tilde{m}_i+\beta q \lambda C^{q-1} m^{q-1}_{i} =0\ .
\eea
\ees
Inserting these into the partition function, we obtain
\bea
Z&=&
\exp\left(- \beta (p-1)J N C^{p} m^{p}  - \beta (q-1)\lambda  C^{q}\sum_{i}  m^{q}_i \right)  \cr
&\times&
 \left(
\prod_{i} (e^{\beta \sqrt{(p J C^{p-1} m^{p-1} + q \lambda C^{q-1} m_i^{q-1})^2+\Gamma^2}}+
e^{-\beta \sqrt{(p J C^{p-1} m^{p-1} + q \lambda C^{q-1} m_i^{q-1})^2+\Gamma^2}}
)
\right)^{C}\ .
\eea
The free energy $F$ defined by $Z=\exp(-\beta N F)$ is
\bea
\label{eq:F-gen}
F&=&(p-1)J C^{p} m^{p}  + (q-1)\lambda C^{q} {1\over N}   \sum_{i}  m^{q}_i   \cr
&-&{C \over \beta N }\log \left[
\prod_{i} \left(e^{\beta \sqrt{(p J C^{p-1}m + q \lambda C^{q-1}m_i)^2+\Gamma^2}}+
e^{-\beta \sqrt{(p J C^{p-1}m + q \lambda C^{q-1}m_i)^2+\Gamma^2}}\right)
\right]\ .
\eea


%
\twocolumngrid

\section{Excited states}
\label{Appendix excited states}

\begin{figure*}[t]
\subfigure[]{\includegraphics[width=0.22\textwidth]{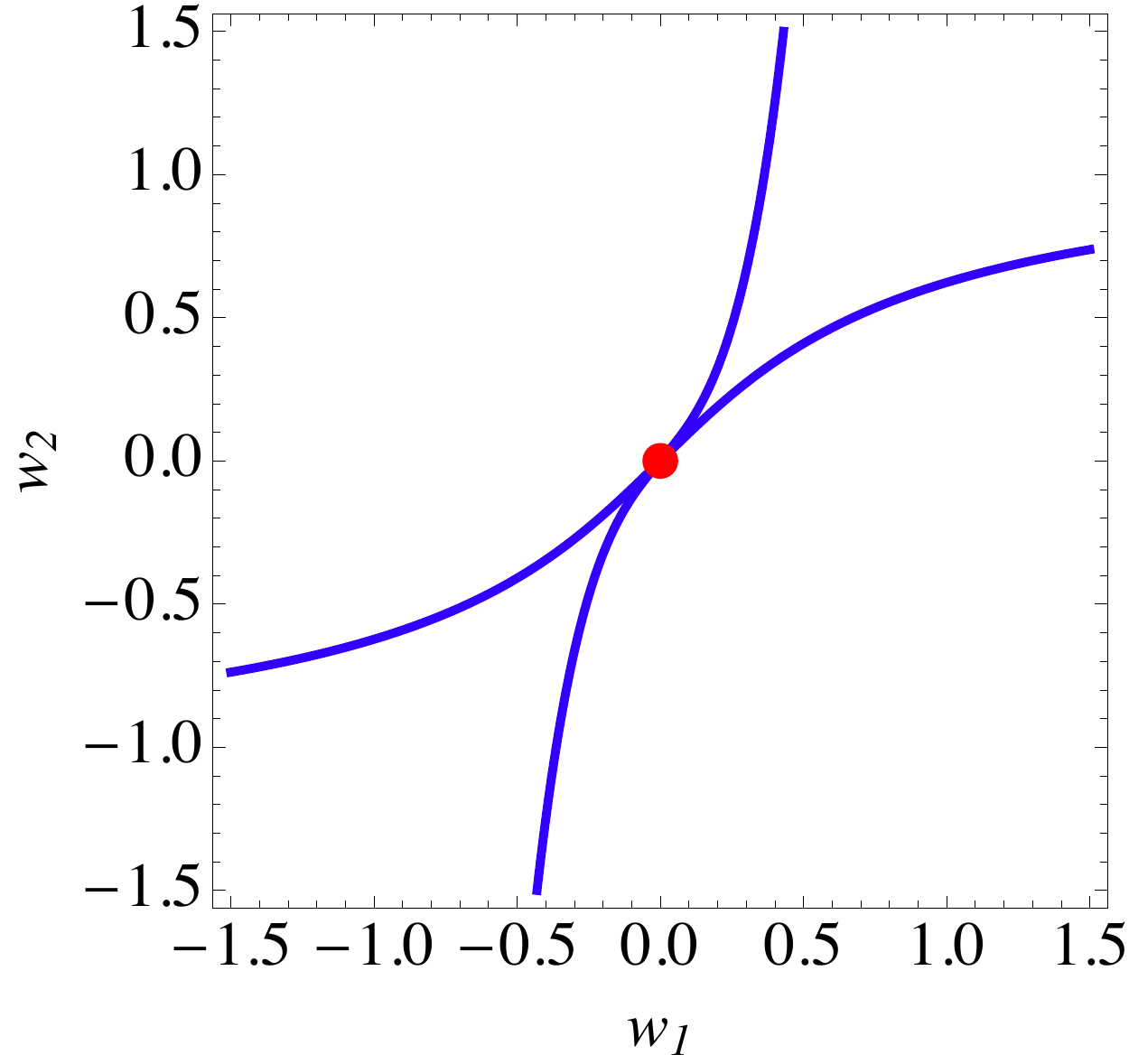}  \label{intersecg09G115}  }
\subfigure[]{\includegraphics[width=0.22\textwidth]{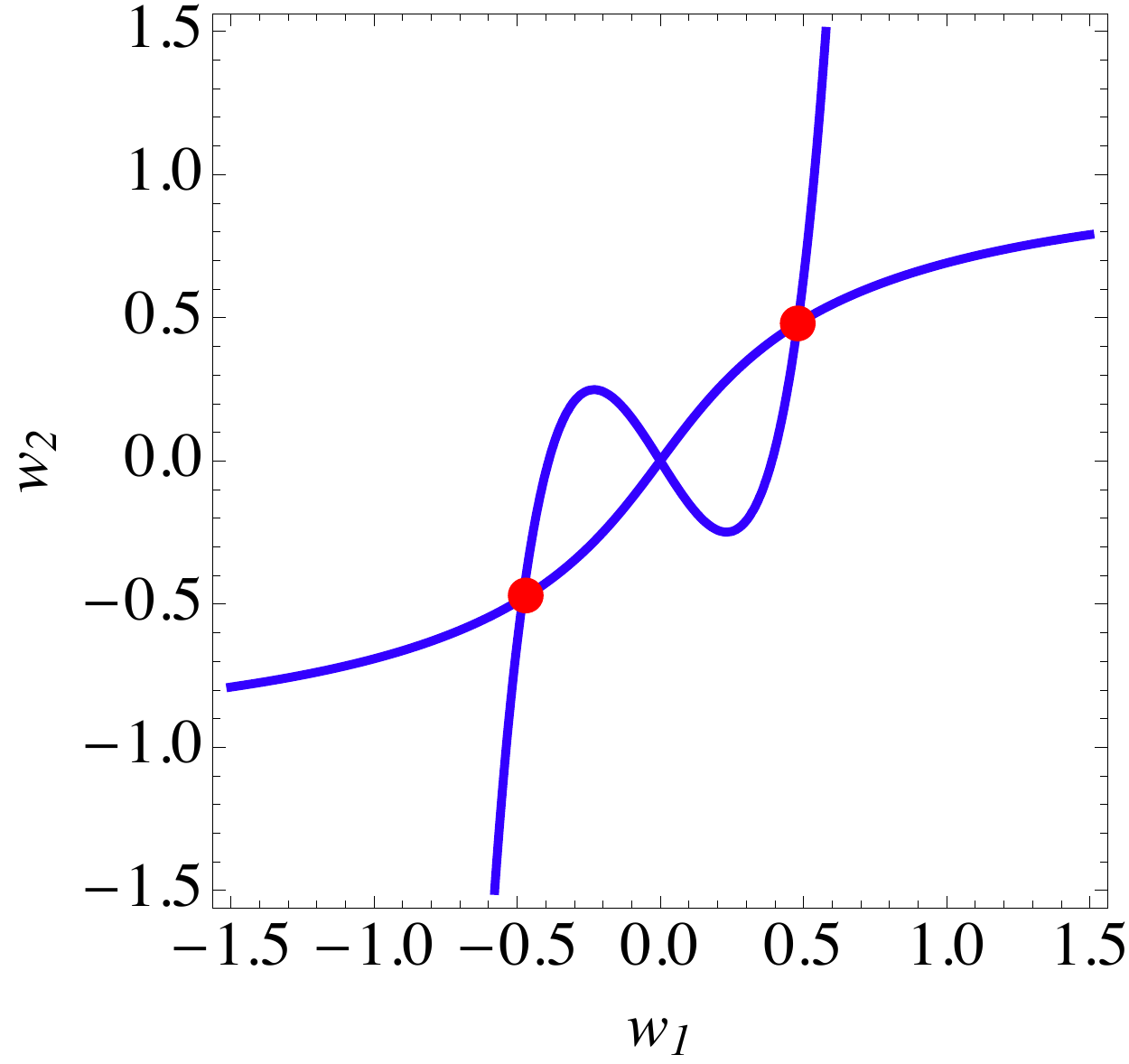}  \label{intersecg09G100}  } 
\subfigure[]{\includegraphics[width=0.22\textwidth]{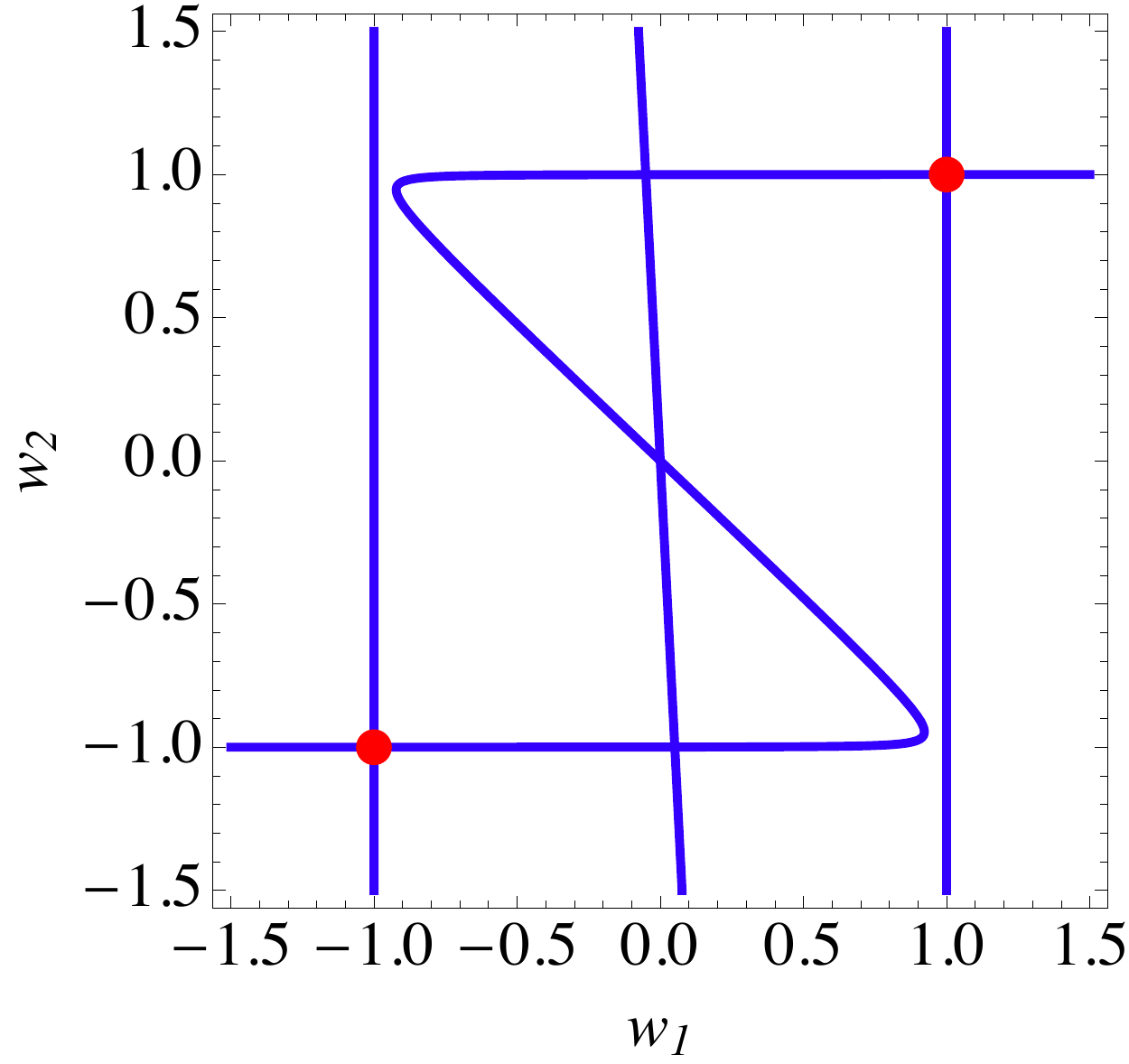}  \label{intersecg09G2}  }
\subfigure[]{\includegraphics[width=0.22\textwidth]{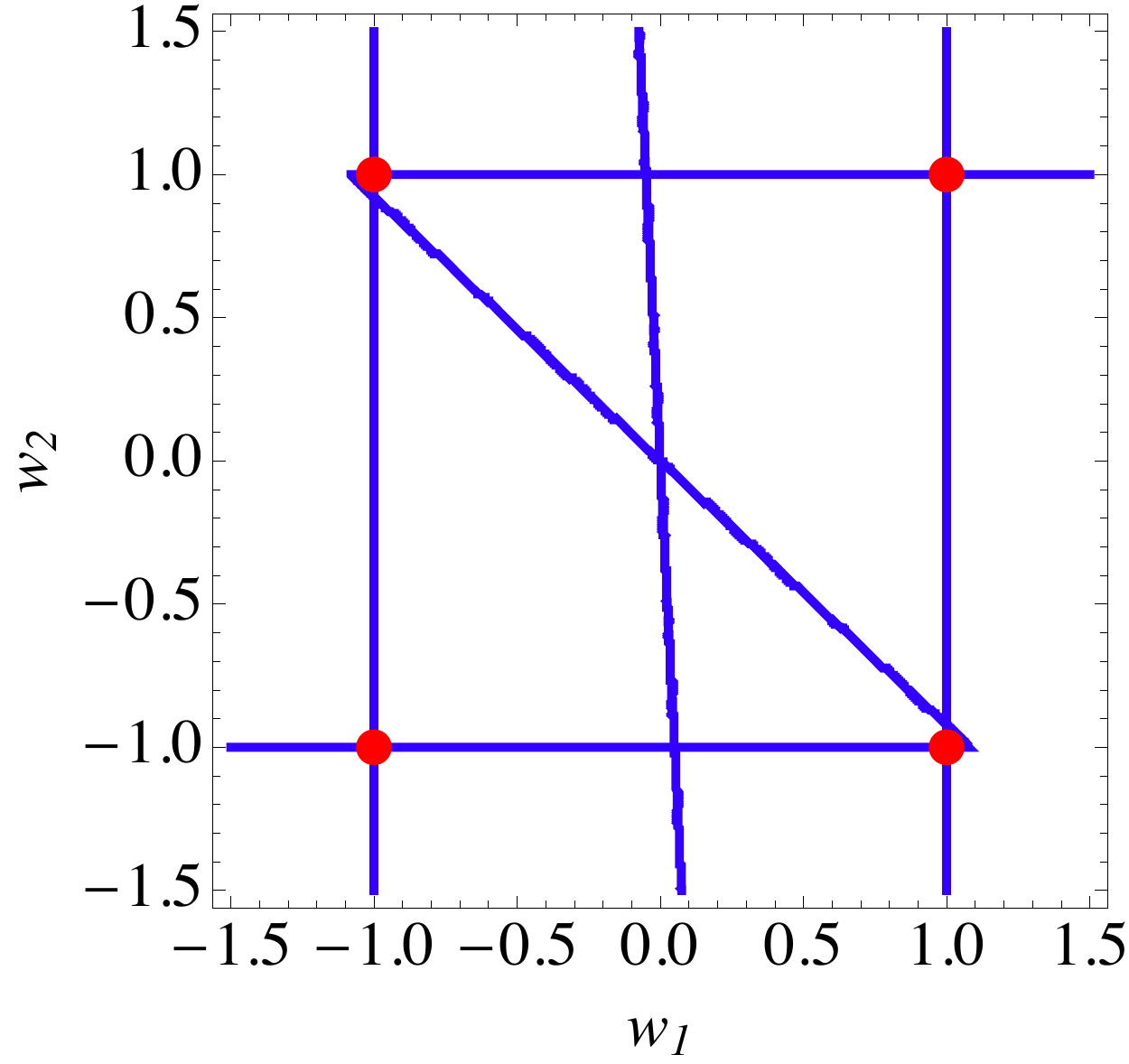}  \label{intersecg09G0}  }
\caption{ 
Solutions to Eqs.~\eqref{ferro-excited cond 1a} and ~\eqref{ferro-excited cond 1b} for 
$k/N=0.09,J=1,T/C=0,\gamma=0.9$.
Red dots are local and global minima.
(a) $\Gamma/C=3.8$;
(b) $\Gamma/C=3.3$;
(c) $\Gamma/C=0.07$;
(d) $\Gamma/C=0$.
}
\label{intersecg09G}
\end{figure*}

In the main text we analyzed ground state phase transitions, where
the main focus was the energy gap between the ground and first excited states.
We demonstrated that this gap becomes larger at the phase transition as the penalty strength increases. When this is the case it is more likely for the system to stay in the ground state, and errors are suppressed.
However, in an open system, thermal fluctuations can cause transitions to excited states 
even when there is no phase transition.
Therefore it is important to understand how easily excited states can relax into the ground state.
The aim of this section is to understand the energy landscape of NQAC in the presence of a strong penalty coupling.

\begin{figure*}[t]
\subfigure[]{\includegraphics[width=0.45\textwidth]{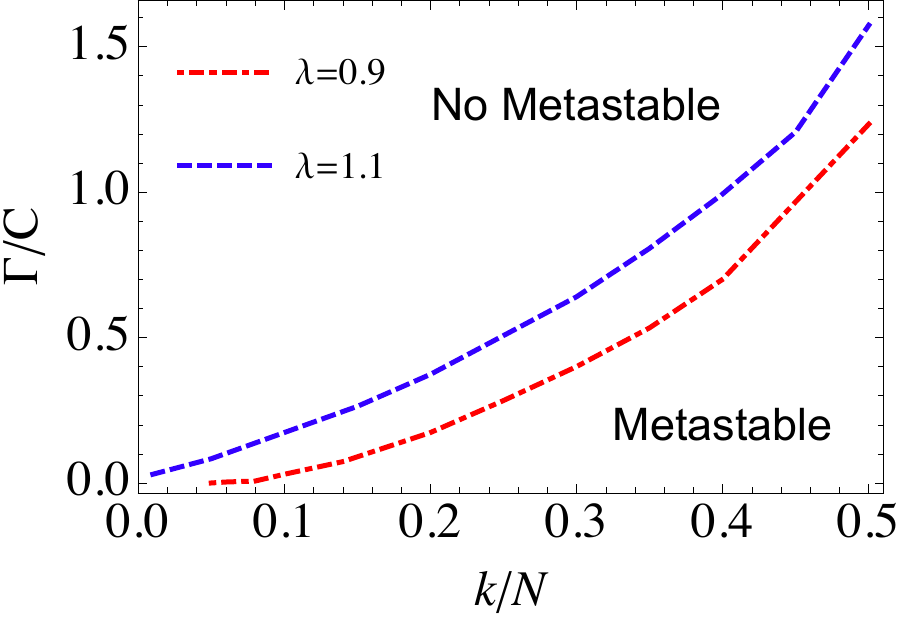}   \label{allptransT0} } 
\subfigure[]{\includegraphics[width=0.45\textwidth]{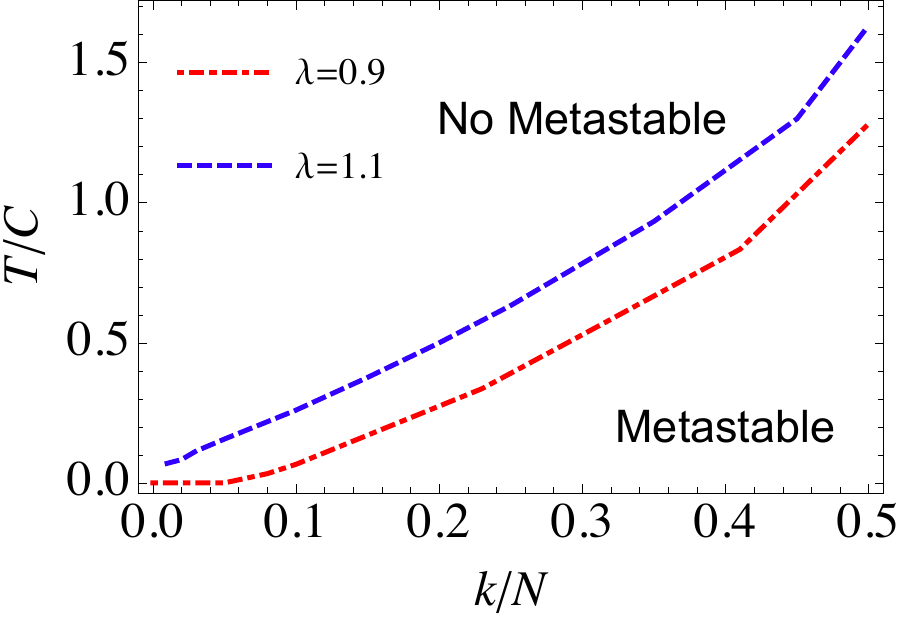}  \label{allptransG0}  } 
\caption{
Phase diagrams in the $k$-$\Gamma$ plane and $k$-$T$ planes. (a): $k/N$ vs $\Gamma/C$ {at $T=0$}. for $\gamma=0.9$ and $\gamma=1.1$, the phase transition line is determined only by the ratio $k/N$:
$k/N = 0.05$ is the lowest excited state which can have a metastable state {for $\gamma=0.9$}. 
(b) $k/N$ vs $T/C$ {at $\Gamma=0$}. for $\gamma=0.9$ and $\gamma=1.1$. the phase transition line is determined only by the ratio $k/N$. 
{$T/C=1.28$ is the highest temperature which can have a metastable state for $\gamma=0.9$}. Above this, metastable state disappears.}
\end{figure*}

\subsection{The ferromagnetic case for $p=2$}
Let us first consider the ferromagnetic case with $p=q=2$.
In the ground state, all the spins take the same expectation value $m_i \equiv w_1$.
The excited states include spin flips within a logical qubit. 
If the number of spin flips is less than $C/2$, we obtain a correct answer by majority vote at the end of annealing.
For simplicity, we do not consider spin flips within a logical qubit and assume that all the encoded qubits in a logical qubit point in the same direction (this would be the case when the penalty coupling is large). 
In the following, we study when they become metastable states (local minima of the free energy).

Without loss of generality, we can assume that $w_1$ is positive and 
$w_2$ is negative.
Substituting this into Eq.~\eqref{ferro-saddle-b}, the excited state is determined by:
\bes
\label{ferro-excited cond 1}
\begin{align}
w_1&={(2J m +2\lambda w_1)\over \sqrt{(2J m+2\lambda w_1)^2+(\Gamma/C)^2}} \times \notag \\
&\tanh(\beta C \sqrt{(2Jm+2\lambda w_1)^2+(\Gamma/C)^2}), \label{ferro-excited cond 1a}\\
w_2&={(2J m +2\lambda w_2 )\over \sqrt{(2Jm+2\lambda w_2)^2+(\Gamma/C)^2}}  \times \notag \\&\tanh(\beta C \sqrt{(2J m+2\lambda w_2)^2+(\Gamma/C)^2}) \label{ferro-excited cond 1b}
\end{align}
\ees
where
\bea
m={(N-k) w_1+ k w_2\over N}.
\label{ferro-excited cond 2}
\eea
The solutions of these equations are shown in Fig.~\ref{intersecg09G}.
The two curves show the individual solutions of Eqs.~\eqref{ferro-excited cond 1a} and 
~\eqref{ferro-excited cond 1b}, and the intersections are the solutions that satisfy both equations.  
The red dots in Fig.~\ref{intersecg09G} are local and global minima. 
When quantum fluctuations are large the only local minimum is the ground state [Fig.~\ref{intersecg09G115}].
Each logical spin is in the paramagnetic state $m_i=0$.
As the fluctuations decrease, there is a second order phase transition and the local minima are in the ferromagnetic states
$m_i\neq 0$ [Fig.~\ref{intersecg09G100}]. 
The value of each logical spin approaches $\pm1$ [Fig.~\ref{intersecg09G2}] and
when the fluctuations become very small, the excited states start to appear as local minima in the mean field free energy [Fig.~\ref{intersecg09G0}].

In the low temperature limit, Eqs.~\eqref{ferro-excited cond 1} become
\bes
\label{ferro-excited condition}
\begin{align}
w_1&={(2J m +2\lambda w_1)\over \sqrt{(2J m+2\lambda w_1)^2+(\Gamma/C)^2}} \\
w_2&={(2J m +2\lambda w_2 )\over \sqrt{(2Jm+2\lambda w_2)^2+(\Gamma/C)^2}} \ .
\end{align}
\ees
The necessary condition for the existence of metastable states is that
 $(w_1,w_2)=(\pm1,\mp1)$ is the solution of Eqs.~\eqref{ferro-excited condition}
 at $\Gamma=0$.
In this limit, Eqs.~\eqref{ferro-excited condition}
become $\text{sgn}(w_{1,2})=\sgn((2J m +2\lambda w_{1,2}))$. 
Then $(w_1,w_2)=(\pm1,\mp1)$ can be a solution if 
\bea
2J-{4Jk\over N}-2\lambda<0
\eea
This condition is satisfied for any value of $(k,N)$ if $J<\lambda$.
If $J>\lambda$, it is satisfied only if the middle term ${4Jk\over N}$ compensates for the difference between $J$ and $\lambda$. Specifically, the metastable state disappears for small $k$ more easily than for large $k$. And for any $k$, the metastable state disappears as $N$ increases.

From this observation, we know that there can exist metastable states when the quantum fluctuations are small. In Fig.~\ref{allptransT0} we show the transition lines in the $(k/N,\Gamma/C)$ plane
below which there exist metastable states and above which there are no metastable states.
This is likewise the case for thermal fluctuations; in Fig.~\ref{allptransG0}, we show the 
line between the existence and non-existence of metastable states
in the $(k/N,T/C)$ plane
Remarkably, Fig.~\ref{allptransG0} and Fig.~\ref{allptransT0} are indistinguishable, suggesting that the excited states are similarly affected by quantum and thermal fluctuations in this case.

%
\begin{figure*}[t]
\subfigure[]{\includegraphics[width=0.45\textwidth]{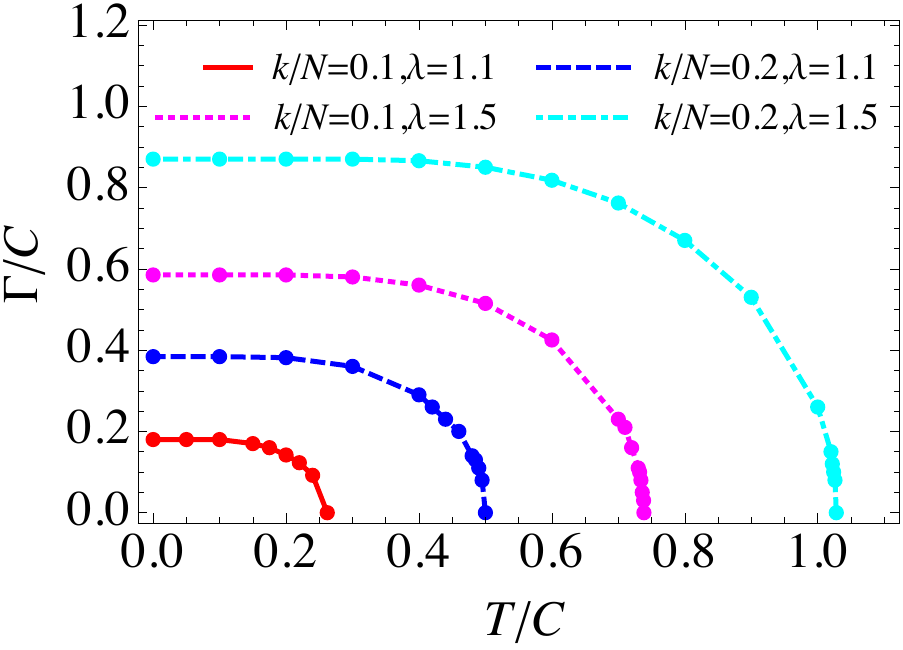}  \label{FexcitedTG}}
\subfigure[]{\includegraphics[width=0.45\textwidth]{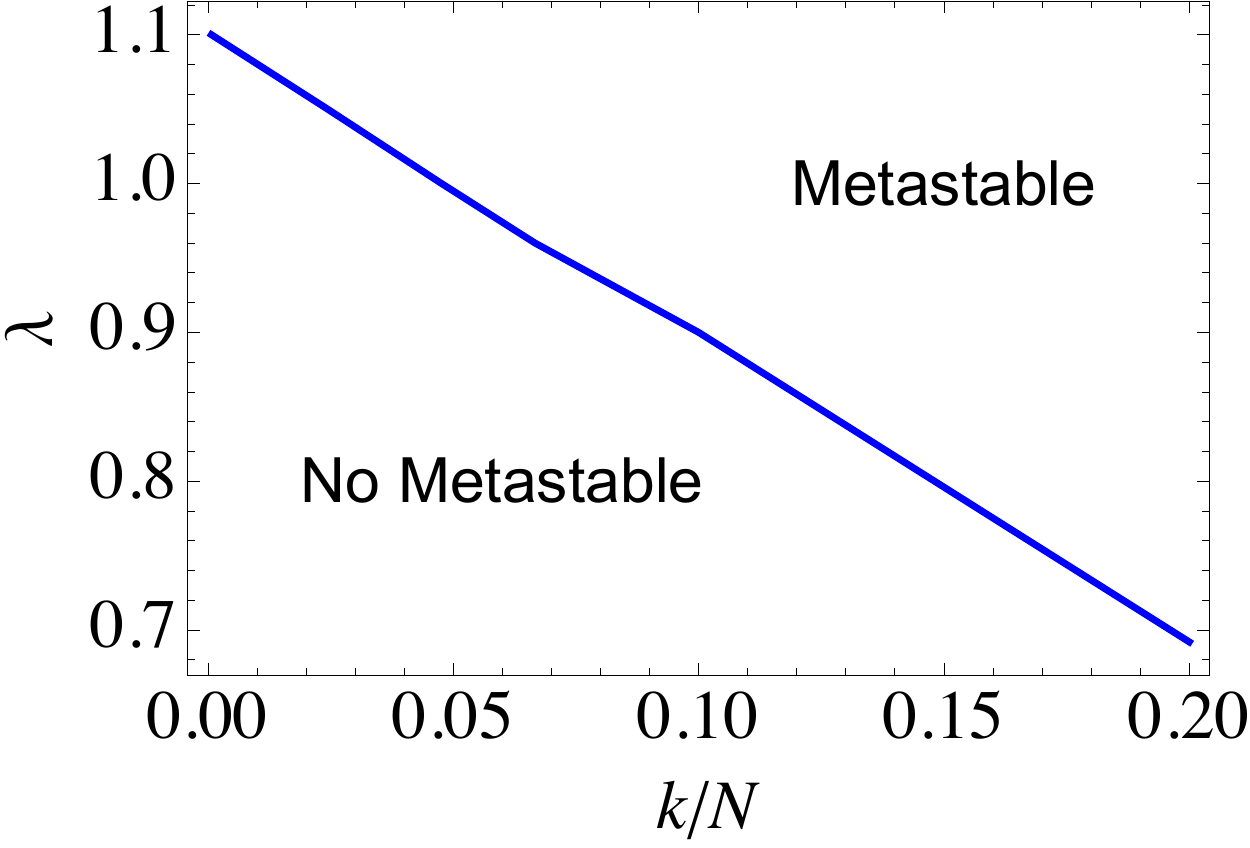}  \label{Fpngb10G01}}
\caption{(a) Disappearance of lower excited states for $k/N=0.1,0.2$ in the ferromagnetic case for two values of $\lambda$: metastable states exist only below the transition lines (toward the bottom left). (b) Critical $\lambda$ as a function of $k/N$ in the ferromagnetic case. The metastable states are first excited states. Here
$T/C=0.03$, $\Gamma/C=0.03$, and $J=1$.}
\label{FexcitedTG-and-Fpngb10G01}
\end{figure*}

Fig.~\ref{FexcitedTG} shows the transition line between the existence and the absence of metastable excited states in the $(T/C,\Gamma/C)$
plane, for various values of $k/N$ and $\lambda$. Fig.~\ref{Fpngb10G01} shows the transition line between the existence and the absence of metastable excited states in the $(k/N,\lambda)$ plane.
For a fixed penalty, when only a small number of spins inside a logical qubit have flipped there are no metastable states. But a state with many flipped spins (at fixed penalty) can be metastable. The higher the penalty, the smaller the range of no metastability, since in the limit of large penalty the only stable states are the all-up and all-down states (logical spins must be completely ordered). This explains the negative slope in Fig.~\ref{Fpngb10G01}.


\subsection{Energy and degeneracy for $p=2$ in the ferromagnetic case}

The existence of metastable states can affect the success probability of quantum annealing due to thermal and quantum fluctuations.
At the end of annealing ($\Gamma=0$), after substituting Eqs.~\eqref{ferro-excited condition} into $H_Z$ [Eq.~\eqref{eq:HZ}], the energy levels are:
\begin{align}
E_k&\simeq C^2\Big[
J  \left({N-k\over N}w_1+{k\over N} w_2\right)^2   \notag \\
&+{\lambda }\left({N-k\over N}w_1^2+{k\over N} w_2^2\right)\notag \\
& -{N-k\over  N}\sum_{i=1}^{N}2\left|J \left({N-k\over N}w_1+{k\over N} w_2\right)+\lambda  w_1\right| \notag\\
&-{k\over  N}\sum_{i=1}^{N}2\left|J \left({N-k\over N}w_1+{k\over N} w_2\right)+\lambda  w_2\right|
\Big]\ .
\end{align}
In thermal equilibrium, the probability $P_{k}$ of finding a state at $E_k$ is
\bea
P_k\sim d_k e^{-\beta E_k}\ ,
\eea
where $d_k$ is the degeneracy. In our case
$d_k=2{N \choose k}$.
For large $N$ and small $k$ we have
$d_k\simeq {(n/k-1/2)^{k}e^{k}\over \sqrt{\pi k/2}}$.
Note that the degeneracy increases rapidly as the energy increases.
Without this entropy effect, the excited states are suppressed by the Boltzmann factor.
However, the rapid increase of $d_k$ for excited states might change this situation and therefore change the phase transition and the annealing success probability. To analyze this,
let us use Eq.~\eqref{eq:F-gen} to define an effective action $F_{k}$ as the free energy of the $k$-th excited states:
\bea
&&F_{k}=JC^2m^2+
{\lambda C^2\over N}((N-k)w_{1}^2+k w_{2}^2) \notag\\
&&-{C\over \beta N} 
\left(
(N-k)\log \left[2\cosh\left(\beta\sqrt{4C^2(Jm+\lambda w_{1})^2+\Gamma^2} \right)\right]  \right.\cr
&&\left.
+k\log \left[2\cosh\left(\beta\sqrt{4C^2(Jm+\lambda w_{2})^2+\Gamma^2} \right)\right]
\right) \notag \\
&&-{1\over \beta N}\log \left[2{N\choose k}\right] ,
\label{free energy with degeneracy}
\eea
which takes into account the entropy (degeneracy) through the last term. The partition function is given by $Z=\sum_{k}e^{-\beta N F_{k}}$.

\begin{figure*}[t]
\subfigure[]{\includegraphics[width=0.45\textwidth]{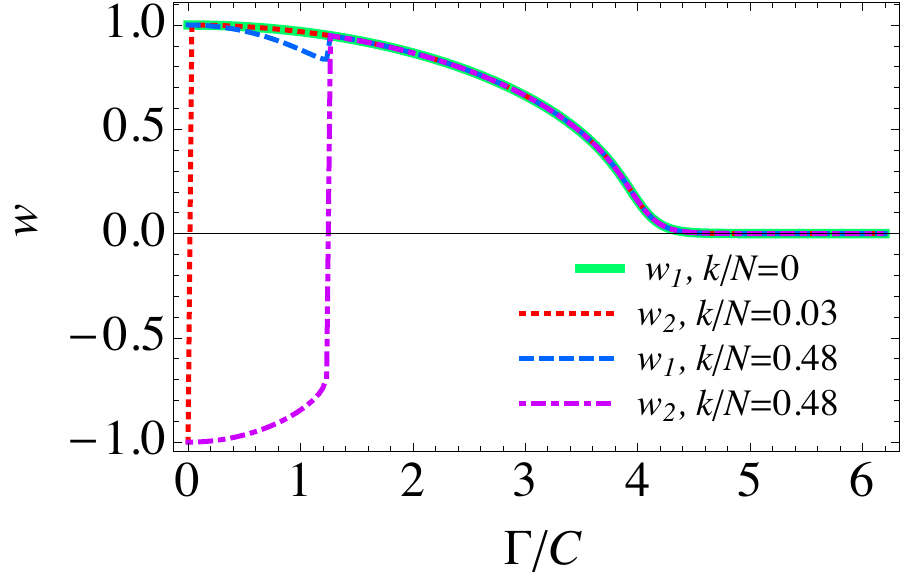} \label{w1w2G} }
\subfigure[]{\includegraphics[width=0.45\textwidth]{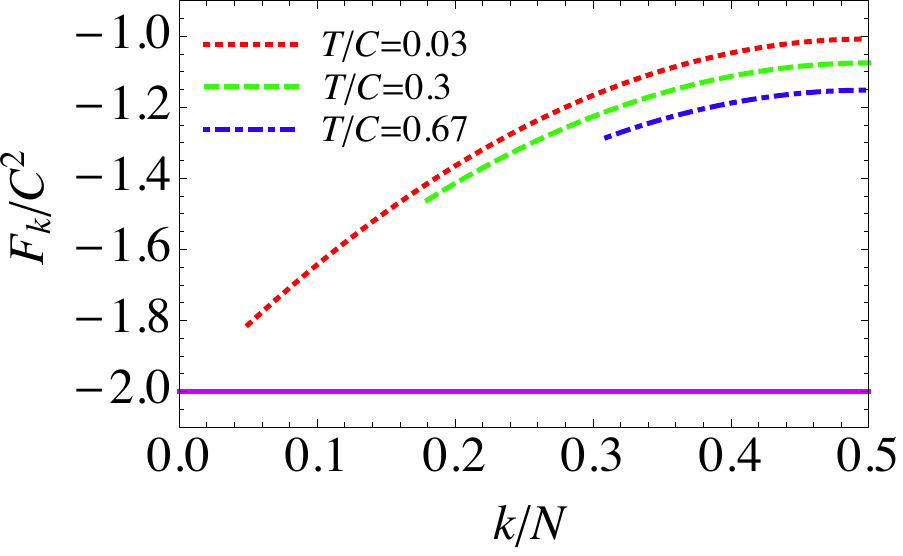}   \label{indivfree} }
\caption{(a) $w_1$ and $w_2$ for the ground state $k/N=0$, $k/N=0.03$ and highly excited state $k/N=0.48$,
as a function of $\Gamma/C$ at $T/C=0.03$, $\lambda=1$. (b) $F_k$ for $\Gamma/C=0.03$.
The solid purple line represents the ground state free energy contribution $F_0/C^2$.}
\end{figure*}

By solving Eq.~\eqref{ferro-excited condition}, we plot in Fig.~\ref{w1w2G} the values of $w_1$ and $w_2$ for the ground state $k=0$,
a slightly excited state $k/N=0.03$, and a highly excited state $k/N=0.48$ at $T/C=0.03$.
The metastable state $w_2\simeq -1$ disappears quickly when $\Gamma/C$ increases from zero to finite value
and the only true ground state $w_1=w_2\simeq 1$ exists.
For highly excited states, $w_1>0$ and $w_2<0$ approach each other as $\Gamma/C$ increases
(the blue dashed and the purple dot-dashed lines) until $\Gamma/C\sim 1.25$ and 
then the metastable states disappear above this value.
In Fig.~\ref{indivfree}, we plot the free energy contribution from excited states at different temperatures. 
Because of the absence of the metastable states for smaller $k/N$, the free energy contribution of the excited states starts from a critical value of $k/N$ for a given temperature.  
The degeneracy contribution to Eq.~\eqref{free energy with degeneracy} is $-{1\over \beta N}\log \left[2{N\choose k}\right]$. There is a unique ground state for $k=0$, for which this term vanishes. 
The largest degeneracy is realized at $k=N/2$. For $k=a N \gg 1, (a\in(0,1/2])$, this term scales as $T\left[ a\log a +(1-a)\log (1-a)\right]$. 
Figure~\ref{indivfree} shows that although the number of degenerate states increases exponentially, the contribution from the higher excited states is still subdominant as long as the 
effective temperature $T/C$ is low. 
Since this term does not depend on $C$, its effect in the normalized free energy $F_k/C^2$ is diminished as $C$ increases. This is another benefit of using NQAC.


%
\begin{figure*}[t]
\subfigure[]{ \includegraphics[width=0.45\textwidth]{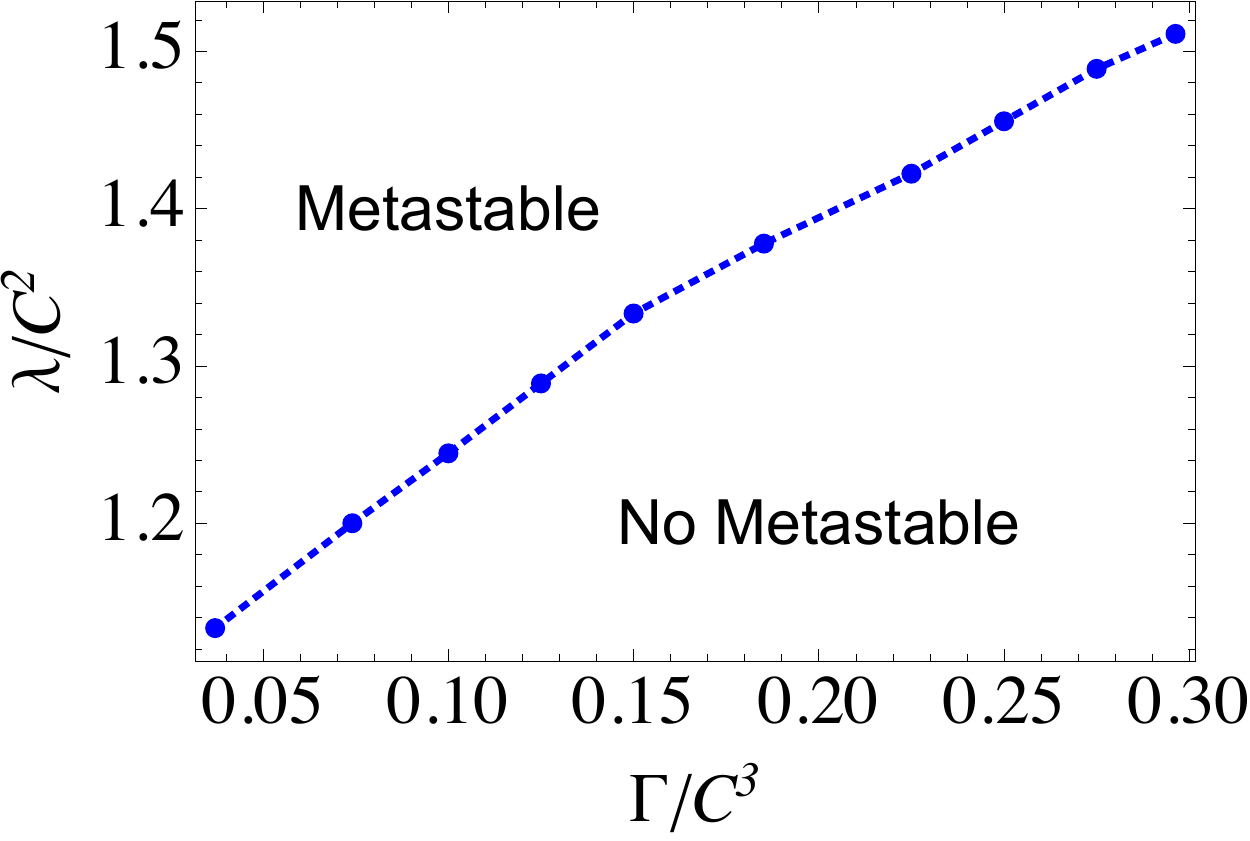} }
\subfigure[]{   \includegraphics[width=0.45\textwidth]{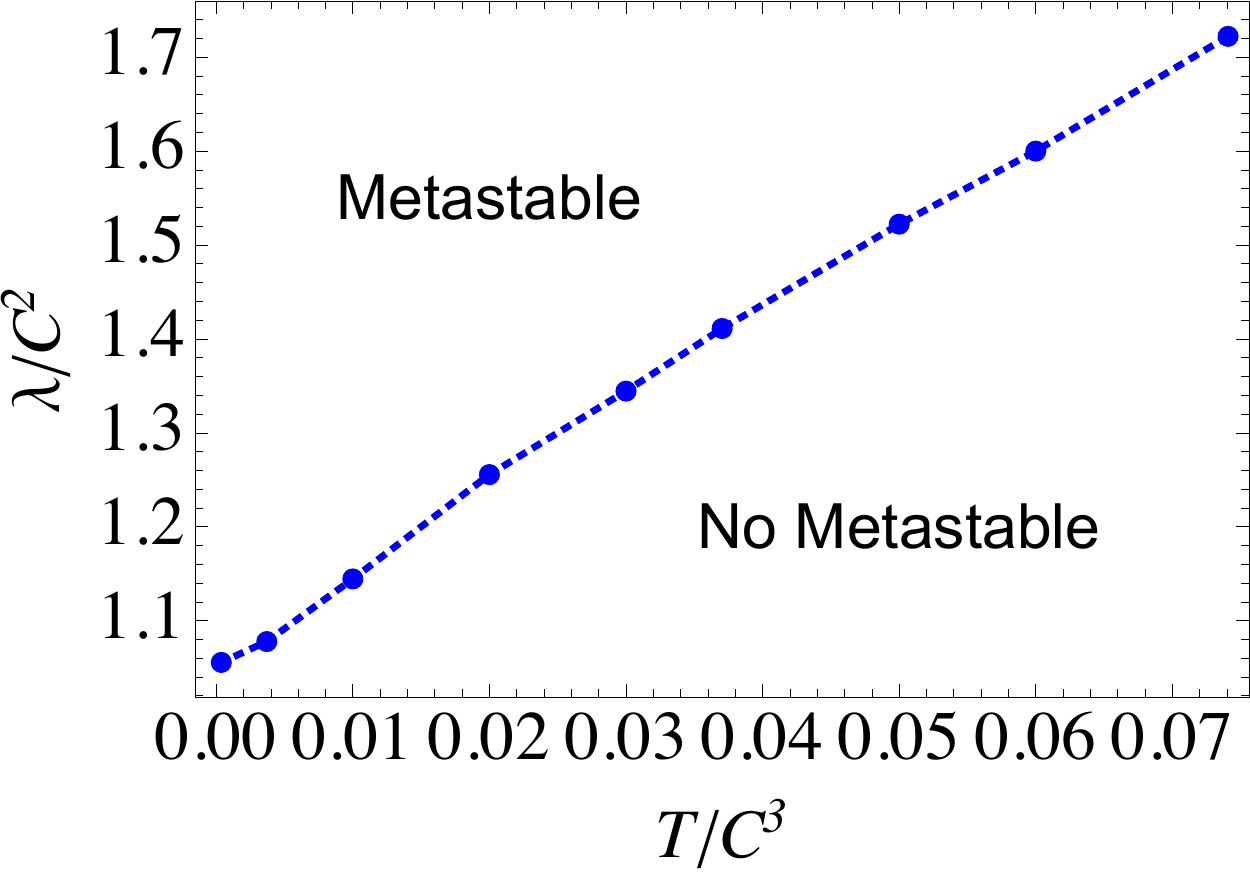}   }
\caption{(a) The borders between the existence and the absence of the metastable states for $k/N=0.1$ and $p=4$. 
(b) Penalty coupling $\lambda/C^2$ as a function of temperature $T/C^3$ for $\Gamma=0$ and a function of $\Gamma$ for $T/C^3=4\times 10^4$ and $p=4$.}
\label{tgammaG0k1}
\end{figure*}

\subsection{Excited states and first order phase transitions when $p\ge3$ in the ferromagnetic case}

\begin{figure*}[t]
 \subfigure[]{\includegraphics[width=0.45\textwidth]{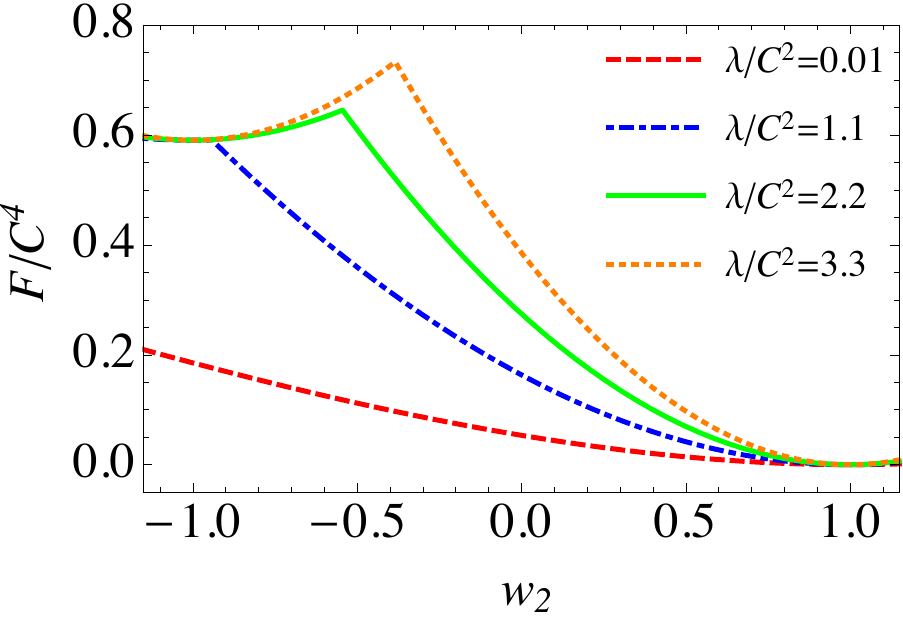}  \label{freek1b100G01} }
 \subfigure[]{   \includegraphics[width=0.45\textwidth]{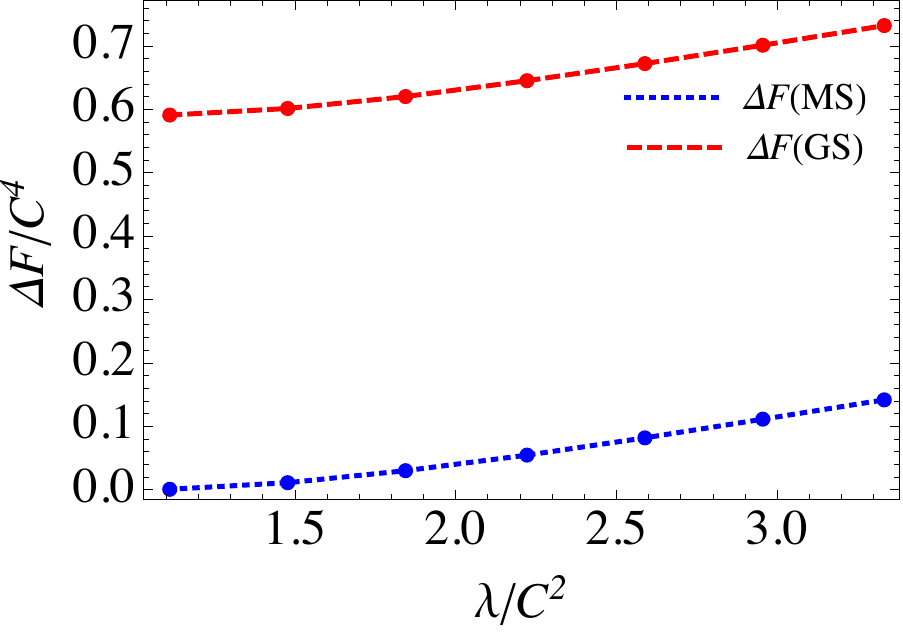}  \label{heightdef}  }
\caption{(a) Normalized free energy $F(w_1=1,w_2)-F(w_1=1,w_2=1)$, for $k/N=0.1$, $T/C^3=3.7\times10^{-4}$, $\Gamma=0.1$. The potential barrier between the  metastable state $(w_1,w_2)\simeq(-1,1)$ and the true ground state
$(w_1,w_2)=(1,1)$ becomes larger as $\lambda/C^2$ increases.
(b) The free energy difference $\Delta F(MS)=F(1,w_2^{max})-F(1,-1)$, $\Delta F(GS)=F(1,w_2^{max})-F(1,1)$ between the local maximum $(w_2=w_2^{max})$, and the local minimum $(w_2=-1)$ or the ground state  $(w_2=1)$  for $w_1=1$ seen in Fig.~\ref{freek1b100G01} (normalized by $C^4$). 
Other parameters:
$k/N=0.1$ $T/C^3=3.7\times 10^{-4}$, $\Gamma/C^3=3.7\times 10^{-3}$.}
\end{figure*}

We can extend the  $p=2$ analysis to $p\ge3$.
Figure~\ref{tgammaG0k1} shows the borders between the existence and absence of the metastable states for $p=4$.
The qualitative features are the same as in the $p=2$ case:  metastable states do not survive at high temperature and/or quantum fluctuations.
Fig.~\ref{freek1b100G01} shows the potential barrier between local minima for various values of the penalty coupling.
The potential barrier becomes higher as the penalty coupling increases.
As one can see in Fig.~\ref{freek1b100G01},
the local free energy minimum is realized at $w_2\simeq -1$ while the local maximum is realized 
at $-1<w_2<0$. On the other hand, the difference between the free energy minima at $w_2\simeq-1$
and $w_2\simeq 1$ does not change as the penalty $\lambda$ changes.

To evaluate the height of the potential barrier, in Fig.~\ref{heightdef} we plot the free energy difference between the local maximum and the local minimum, as well as the local maximum and the ground state seen in Fig.~\ref{freek1b100G01}. 
It shows that increasing the penalty $\lambda$ prevents the ground state to transit into the metastable state.


\section{Excited states in the antiferromagnetic $p=2$ case}

For simplicity let us assume that $N$ is an even number. Classically, the ground state of an antiferromagnetic system then has half of the spins pointing in the positive $z$ direction and the rest of the spins pointing in the negative $z$ direction: the ground state has $N/2$ spins with $m_i=1$ and 
$N/2$ spins with $m_i=-1$, and the average is $m=0$.
The $k$-th excited states have $N/2\pm k$ spins with $m_i=1$ and $N/2\mp k$ spins with $m_i=-1$. The average is $m=\pm{2k\over N}$.
When quantum and thermal fluctuations are induced ($\Gamma, T\neq 0$), the values of $m_i$ change. 
We study the existence and stability of excited states using a mean field analysis.

Consider the $k$-th level excited states in which ${N/2}-k$ of the $m_i$ are positive and ${N/2}+k$ of the $m_i$ are negative (since the system has $Z_2$ spin-flip symmetry we can make this choice without loss of generality).
Similarly to Eq.~\eqref{ferro-excited cond 1}, to determine the excited states we need to consider the following set of equations:
\bes
\label{ferro-excited cond 1-check}
\begin{align}
w_1&={(-2J m +2\lambda w_1)\over \sqrt{(-2J m+2\lambda w_1)^2+(\Gamma/C)^2}}\times  \notag \\
&\tanh(\beta C \sqrt{(-2Jm+2\lambda w_1)^2+(\Gamma/C)^2})\\
w_2&={(-2J m +2\lambda w_2 )\over \sqrt{(-2Jm+2\lambda w_2)^2+(\Gamma/C)^2}}\times  \notag \\
&\tanh(\beta C \sqrt{(-2J m+2\lambda w_2)^2+(\Gamma/C)^2}) \\
w_3&={(-2J m +2\lambda w_3)\over \sqrt{(-2J m+2\lambda w_3)^2+(\Gamma/C)^2}}\times \notag \\
&\tanh(\beta C \sqrt{(-2Jm+2\lambda w_3)^2+(\Gamma/C)^2})\ ,
\end{align}
\ees
where
\bea
m={({N\over 2}-k-1) w_1+ ({N\over 2}+k) w_2+ w_3\over N}\ .
\label{ferro-excited cond 2-check}
\eea
Here $({N/2}-k-1)$ of the $m_i$ take a value $w_1>0$, $({N/2}+k)$ take a value $w_2<0$, and a single spin takes a value $w_3$, which can be positive or negative.
If $w_3$ is positive, then it is the $k$-th excited state and if it is negative, then it is the $(k+1)$-th excited state.
The reason why $w_3$ is needed is the following, and stems from the the difference between the ferromagnetic and antiferromagnetic cases. In the ferromagnetic case, we considered the transition between the $k$-th excited state and the ground state.
The $k$-th excited state has $N-k$ spins taking a positive value $w_1>0$ and $k$
spins taking a negative value $w_2<0$, while the ground state has all spins taking a positive value. The reason we consider this transition is that lower energy states, for instance the $(k-1)$-th excited state, are less stable than
higher energy states against the quantum and thermal fluctuations. 
This feature can be seen in Fig.~\ref{FexcitedTG-and-Fpngb10G01} as well as Fig.~\ref{w1w2G}.
Therefore, when 
the $k$-th excited state loses its metastability, 
the $(k-1)$-th excited state has already lost its meta-stability, and the transition happens 
collectively with all $k$ spins flipping from negative to positive.
This kind of behavior is captured by two parameters $w_1$
and $w_2$.
On the other hand, in the antiferromagnetic case, the 
lower excited states are more stable; when the $k$-th excited state loses its metastability, all the lower energy states are still metastable.
Therefore, the transition from the $k$-th excited state to the $(k-1)$-th excited state happens most likely by flipping a single spin. This transition requires three parameters $w_{1,2,3}$; two of them ($w_{1}$ and $w_2$) do not change 
their signs before and after the transition, and one
of them ($w_3$) describes the single spin that flips under the transition.

\begin{figure*}[t]
\subfigure[\ $T/C=0$]{   \includegraphics[width=0.45\textwidth]{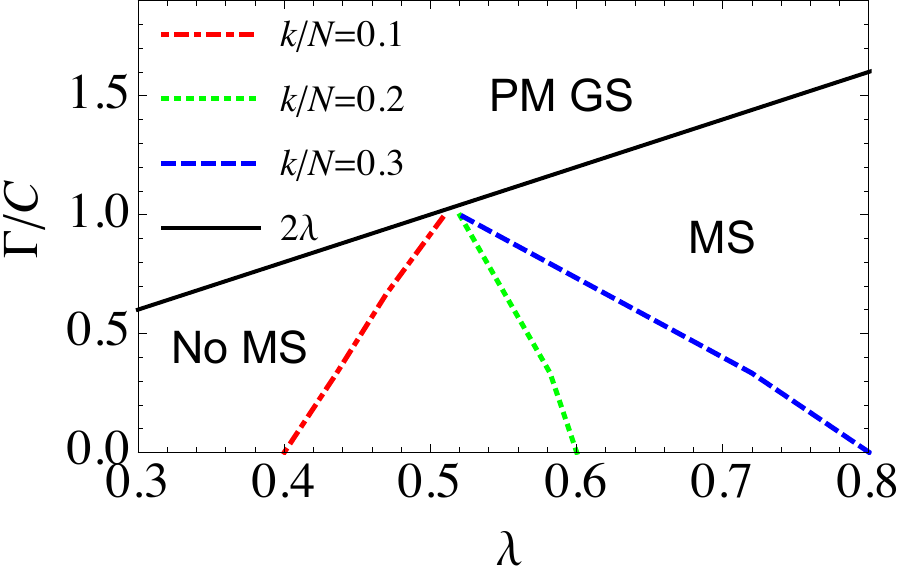} \label{AFphasetransT0} }
\subfigure[\ $T/C=0.33$]{   \includegraphics[width=0.45\textwidth]{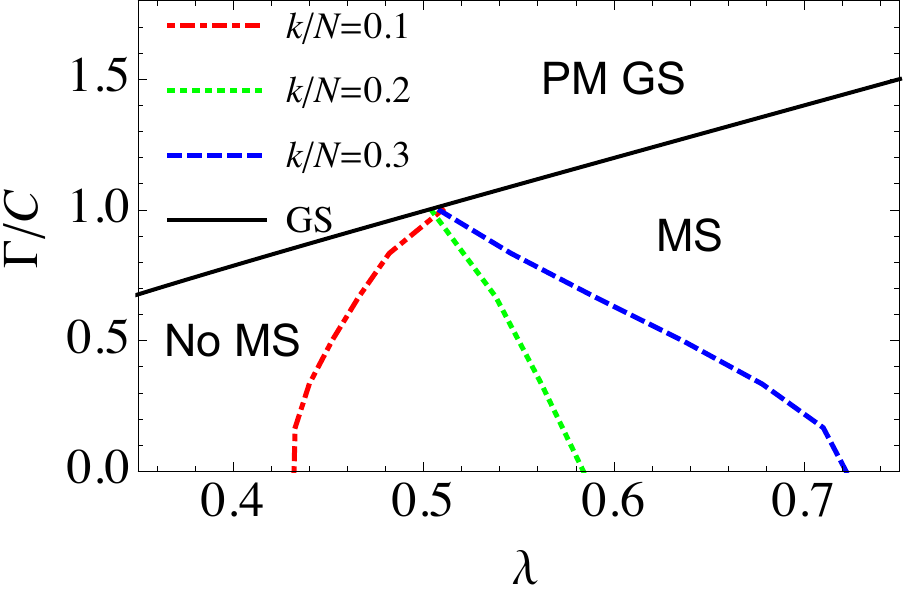} \label{AFphasetransT1}}
\caption{ 
Existence of metastable excited states in the antiferromagnetic case. Panel (a): $T/C=0$; panel (b): $T/C=0.33$.
Red dot-dashed, green dotted, and blue dashed lines represent $k/N=0.1,0.2$ ,and $0.3$, respectively.
To the right of these lines the corresponding excited states exist as metastable states (MS),
while to the left the corresponding excited states do not exist as metastable states (No MS).
Above the black solid line, only the paramagnetic ground state (PM GS) exists and the free energy does not have a local minimum.
  }
\label{AFphasetrans}
\end{figure*}

At $T=0$ and $\Gamma=0,$ the saddle point equations are
\bea
w_1&=&\text{sgn}(-2J m +2\lambda w_1)\cr
w_2&=&\text{sgn}(-2J m +2\lambda w_2 )\cr
w_3&=&\text{sgn}(-2J m +2\lambda w_3)
\label{zero TG saddle eq AF}
\eea
In this case, $w_i$ can only be $\pm1$.
Let us assume $w_1=1$ and $w_2=-1$. For $k>0$, $w_3=+1$ has a lower energy and $w_3=-1$ has a higher energy. In order to have $w_1=1$, $w_2=-1$, $w_3=1$
as the solution of Eq.~\eqref{zero TG saddle eq AF},
the couplings must satisfy
\bea
\lambda>{2k J\over N}
\label{lower energy cond}
\eea
while 
for $w_1=1$, $w_2=-1$, $w_3=-1$, we need
\bea
\lambda>{2(k+1) J\over N}
\label{higher energy cond}
\eea
As can be seen from Eq.~\eqref{lower energy cond}, the ground state $k=0$ always satisfies the existence condition.
Higher energy states $(k\neq0)$ do not exist unless the penalty coupling $\lambda$ is large enough.

At finite temperature and finite transverse field ({$T\neq 0,\Gamma \neq0$}), we solve the saddle point equations~\eqref{ferro-excited cond 1-check} numerically
for various values of the transverse field $\Gamma$ and the penalty $\lambda$.
The dashed lines in Fig.~\ref{AFphasetrans} separate the different phases. The excited states corresponding to each value of $k/N$ exist as metastable states of 
the free energy in the larger $\lambda$ region of the transition lines, while there are no corresponding metastable states in the smaller $\lambda$ region.
The solid line represents a phase transition line in the ground state. The ground state is locally paramagnetic ($m_i=0$) above the line, and $m_i\neq0$ below the line.
For small excitations $k/N \ll 1$ (red dotted lines in Fig.~\ref{AFphasetrans}) the critical $\Gamma$ increases as $\lambda$ increases, while for larger $k/N$ (green and blue dotted lines in Fig.~\ref{AFphasetrans}) the critical $\Gamma$ decreases.
This means that for a given penalty coupling $\lambda$, the  metastable states of higher energy states disappear at larger values of $\Gamma$. Therefore, they are less stable. This result is the opposite of the ferromagnetic case, where higher excited states are locally more stable than lower excited states.

\begin{figure}[t]
 \includegraphics[width=0.45\textwidth]{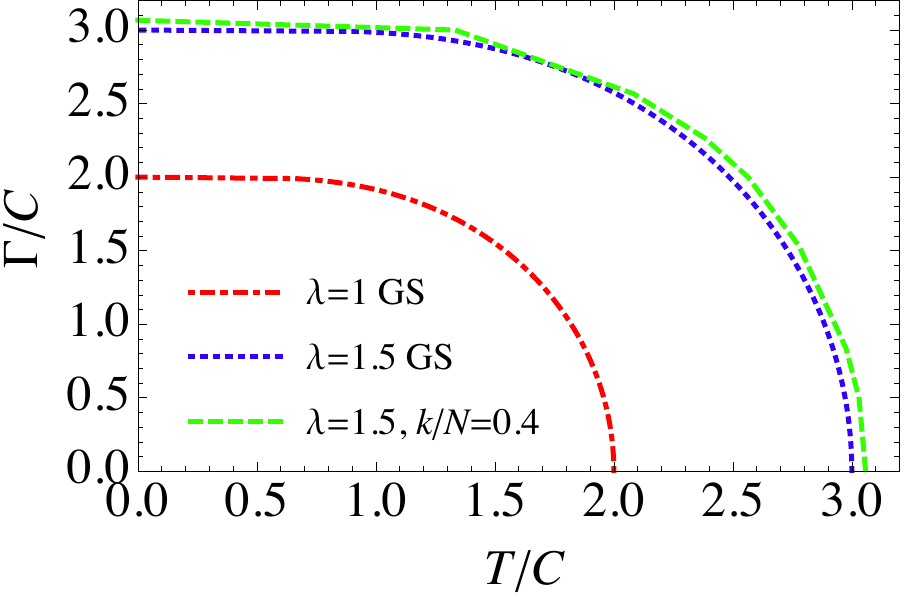}
\caption{ Phase diagram in the $T$-$\Gamma$ plane between the phases with $m=0$ and $m\neq0$, for $\lambda=1.5$ and $\lambda=1$.
}
\label{TvsGg15g1}
\end{figure}
\begin{figure*}[t]
\subfigure[]{\includegraphics[width=0.45\textwidth]{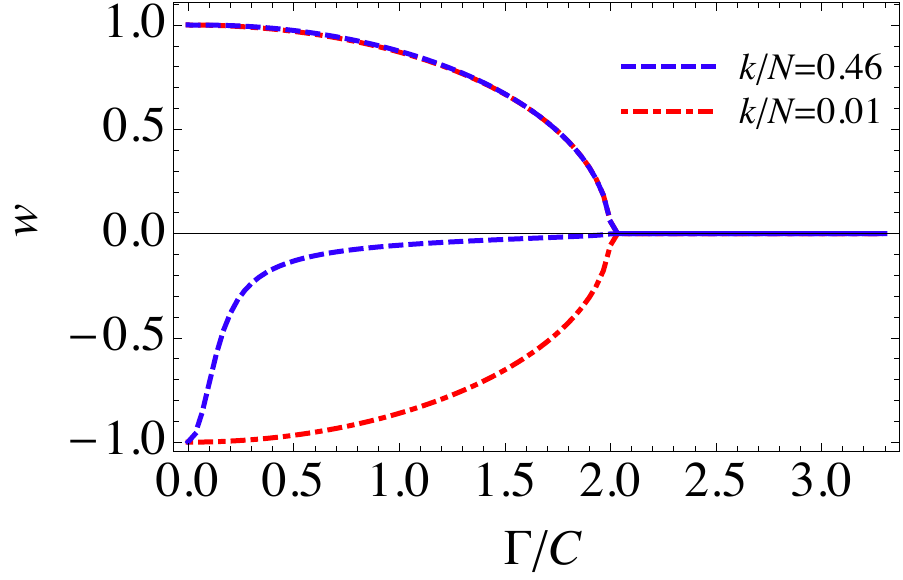}  \label{w1w2b10 w-value AF-a} }
\subfigure[]{\includegraphics[width=0.45\textwidth]{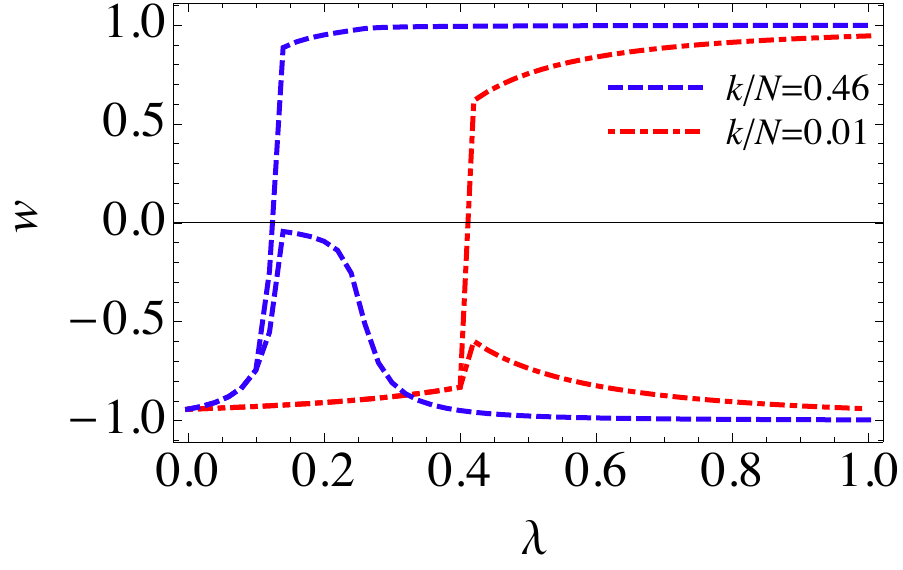} \label{w1w2b10 w-value AF-b}}
\caption{$w_1$ (starts from $1$ at $\Lambda=0$) and $w_2$ (starts from $-1$ at $\Lambda=0$) for a low energy state $k/N=0.01$ and highly excited state $k/N=0.46$. Here 
$T/C=0.03$, $\lambda=1$
(b) $w_1$ (ends at $1$ at $\gamma=1.0$) and $w_2$ (ends at $-1$ at $\gamma=1.0$) for $\Gamma/C=0.67$ and $T/C=0.03$. 
 }
  \label{w1w2b10 w-value AF}
\end{figure*}

Fig.~\ref{TvsGg15g1} shows the phase transition lines in the ($T/C,\Gamma/C$) plane
for the penalty coupling values $\lambda =1,1.5$.
The phase transition for the ground state between the local paramagnetic $m_i\neq0$ and ferromagnetic states $m_i\neq0$ are given by Eq.~\eqref{locally broken condition}.
The green dotted line is for the excited states $k/N=0.4$. One can see that the phase transition lines are fairly insensitive to the excitation level $k/N$.

In the ferromagnetic case, there are two degenerate ground states for locally ordered phase $m_i=\pm n$.
In the antiferromagnetic case, there is an ${N\choose N/2}$-fold degeneracy in the ground state for even $N$
and an ${N\choose (N-1)/2}$-fold degeneracy for odd $N$.
This is in contrast to the case of the one-dimensional Ising model with nearest-neighbor coupling $H=\pm\sum_{i}\sigma^{z}_i\sigma^{z}_{i+1}$.
In this case, there are two ground states for both ferromagnetic and antiferromagnetic interactions.
In the nested configuration, all spins are coupled. Therefore, any configurations
which have the same number of positive $m_i$ and negative $m_i$ have the same energy. 
This fact may change the thermal behavior of the ferromagnetic and the antiferromagnetic cases.
In the case of the ferromagnetic model, the number of excited states (metastable states) increases rapidly and therefore their contributions to the partition function and the transition rates between potential wells are important.
On the other hand, in the antiferromagnetic case, the largest degeneracy happens in the lowest energy states and the degeneracy of the excited states is smaller.

We study the state in which $({N/2}-k)$ spins take values $m_i=w_1$,
and $({N/2}+k)$ spins take values $m_i=w_2$.
The ground state is $k=0$.

In Fig.~\ref{w1w2b10 w-value AF}, we show 
 $w_1$ and $w_2$ for the low energy state $k/N=0.01$ and a highly excited state $k/N=0.46$ for $T/C=0.03$. Both $|w_1|$ and $|w_2|$ decrease 
 continuously as a function of $\Gamma$ and reach zero at a critical value.
This is again a second order phase transition.
It is interesting to note [see Fig.~\ref{w1w2b10 w-value AF-a}] that the positive expectation value $w>0$ (minority of the spins) as a function of $\Gamma/C$ does not strongly depend on $k/N$ 
whereas the negative expectation value $w<0$ (majority of the spins) is sensitive to $k/N$. Also interesting is the fact that [see Fig.~\ref{w1w2b10 w-value AF-b}] for small ($k/N$-dependent) penalty $\lambda$, at first the system tends to become locally disordered (both $w_1$ and $w_2$ decrease in absolute value), before settling in the locally ordered state 

Let us  compare the above result with the results obtained in Ref.~\cite{vinci2015nested}.
Let us assume that $\lambda$ is large enough and all the physical qubits within a logical qubit are aligned.
Under this assumption, 
\bea
m={2\left({N\over 2}-k\right)\over N},~~m_{i}=\pm1 \ ,
\eea
where there are ${N\over 2}-k$ logical qubits with $m_{i}=\pm1$ and 
${N\over 2}+k$ logical qubits with $m_{i}=\mp1$.
The energy  at the end of the anneal $(t=t_f,\Gamma=0)$ is then
\bea
E_{k}=C^2\left[J\left(1-{2k\over N}\right)^2+\lambda \right]\ .
\eea
Therefore the energy level difference between the ground state $k=N/2$ and the
$\Delta k$-th excited state $k=N/2\pm \Delta k$ is
\bea
E_{N/2\pm\Delta k}-E_{N/2}={4C^2J\over N^2}(\Delta k)^2\ .
\eea
The probability of occupancy of the ground state is then
\begin{align}
P(J,C)&={d_0 e^{-\beta NC^2\lambda }\over 
\sum_{i=0}^{N/2}d_i e^{-\beta NC^2\lambda 
+\beta {4C^2J\over N}i^2
}} \notag \\
&=
{1\over 
\sum_{i=0}^{N/2}{d_i\over d_0} e^{
+\beta {4C^2J\over N}i^2
}}\
\end{align}
where $d_i$ is the degeneracy of $i$-th state.
Note that changing the number of physical qubits $C$ is effectively the same as
changing the antiferromagnetic coupling $J$
 \bea
 P(J,C)=P(JC^2,1)
 \eea
This reproduces the theoretical scaling results reported in Ref.~\cite{vinci2015nested}, which were confirmed with empirical data from experiments using a D-Wave processor, with the scaling $J\mapsto JC^2$ replaced by $JC^\alpha$, with $1<\alpha<2$ due to embedding and other overhead.


\section{Energy gap in the second order phase transition}
\label{More on the Energy Gap}

In this section, we calculate the energy gap
following the method described in~\cite{Seoane2012}, which is originally from~\cite{PhysRevA.83.022327}.
The basic idea is to assume $C\gg 1$ and treat $\siz$ and $\six$ semi-classically.
More precisely, we apply the Holstein-Primakoff transformation to the logical qubit
at each site $i$ and treat quantum fluctuations around the classically fixed spin orientation
as small perturbation represented by harmonic oscillators.

It is first necessary to rotate the axes in spin space such that the classical orientation
lies along the new $z$ axis in order to apply the Holstein-Primakoff transformation.
We have in mind the state with vanishing total $z$-magnetization, $S^z=0$,
which is naturally expected to be realized at the beginning and the end of annealing.
Correspondingly, we choose the classical orientation differently for
$i=1,2,\cdots, N/2$ and $i=N/2+1,\cdots, N$,
\begin{equation}
\begin{pmatrix}
\six \\ \siz
\end{pmatrix}
=\begin{pmatrix}
\mp \sta& \cta \\
\cta & \pm \sta
\end{pmatrix}
\begin{pmatrix}
\sixt \\ \sizt
\end{pmatrix}
,
\label{rotation}
\end{equation}
where the upper sign is chosen for $i=1,\cdots, N/2$ and the lower sign for $i=N/2+1,\cdots, N$.
Note that the initial $(\Gamma \to\infty)$ and final $(\Gamma =0)$ orientations
are expected to be $(\six,\siz)=(1,0)$ and $(\six,\siz)=(0,\pm 1)$, respectively,
because the spins align in the $x$ direction initially and along the $\pm z$ direction finally.
This is indeed realized by the above transformation with $\theta =0$ and $\theta =\pi /2$,
respectively, both with $(\sixt,\sizt)=(0,1)$. We now insert the rotation ~\eqref{rotation}) into each term of the Hamiltonian,

\onecolumngrid

\bea
JNC^2 (S^z)^2&=& \frac{JC^2}{N} \left( \sum_{i=1}^{N/2}(\cta\cdot \sixt +\sta\cdot\sizt) 
+\sum_{i=N/2+1}^N (\cta \cdot \sixt -\sta\cdot \sizt) \right)^2 \label{szt} \cr
-JC^2 \lambda \sumi (\siz)^2 &=&
-JC^2\lambda \left( \sum_{i=1}^{N/2}(\cta\cdot \sixt +\sta\cdot\sizt)^2 
+\sum_{i=N/2+1}^N (\cta \cdot \sixt -\sta\cdot \sizt)^2 \right) \label{sizt}\cr
-\Gamma C\sumi \six &=&
-\Gamma C \left( \sum_{i=1}^{N/2}(-\sta\cdot \sixt +\cta\cdot\sizt) 
+\sum_{i=N/2+1}^N (\sta \cdot \sixt +\cta\cdot \sizt) \right). \label{sixt}
\eea

Let us apply the Holstein-Primakoff transformation whose general form for spin of magnitude
$S (\gg 1)$is
\bea
&&S^z=S-a^{\dagger}a,~ \cr
&&S^+=\big(\sqrt{2S-a^{\dagger}a}\big) a \approx \sqrt{2S}\, a,~ \cr
&&S^-=a^{\dagger}\sqrt{2S-a^{\dagger}a}\approx \sqrt{2S}\, a^{\dagger}
\eea
Since the largest possible value of $C\six$ and $C\siz$ is $C$ and also since
$C\sizt$ is supposed to be close to its classically largest value $C$,
we apply this transformation as
\begin{equation}
C\sizt = C-a_i^{\dagger}a_i,\quad C\sixt =\sqrt{\frac{C}{2}} (a_i+a_i^{\dagger}).
\end{equation}
We plug this transformation into Eqs. ~\eqref{szt})-~\eqref{sizt}) and
drop terms beyond harmonic (quadratic) in the boson operators.

The leading-order term of this semi-classical approximation represents
the classical state ($\sizt=1,~\sixt=0$) with energy
\begin{equation}
H_0=-NC\big( JC\lambda \sin^2 \theta +\Gamma \cta\big).
\end{equation}
Minimization of $H_0$ with respect to $\theta$ gives $\theta =0$ or
\begin{equation}
\cos \theta =\frac{\Gamma}{2JC\lambda}. \label{cth}
\end{equation}
The solution $\theta =0$ is to be accepted for $\Gamma >\Gamma_{\rm c}=2JC\lambda$,
whereas the solution to Eq.~\eqref{cth}) is chosen for $0\le \Gamma \le \Gamma_{\rm c}$.
The solution for $\Gamma =0$ is $\theta =\pi /2$ as expected. The next order term is linear in the boson operators, and is expected to vanish
if we choose $\theta$ that minimizes $H_0$.
The reason is that the stationarity condition of the leading term
implies a vanishing linear term. 
 The next term is quadratic in bosons and represents quantum fluctuations around
the classical state.
Diagonalization of this quadratic term gives the energy spectrum, from
which we extract the energy gap.
We therefore drop the terms linear in $a_i^{\dagger}$ and $a_i$ as well as  the
0th order terms as

\begin{align}
 (S^z)^2&=\frac{1}{N^2} \left\{ \sum_{i=1}^{N/2}
\left(\frac{\cta}{\sqrt{2C}}(a_i+\aidag )+\sta \big(1-\frac{\aidag a_i}{C}\big) \right)
+ \sum_{i=N/2+1}^{N}
\left(\frac{\cta}{\sqrt{2C}}(a_i+\aidag)-\sta \big(1-\frac{\aidag a_i}{C}\big) \right)\right\}^2\nonumber\\
&\to \frac{1}{N^2}\frac{\cos^2\theta}{2C}\left(\sumi (a_i+\aidag)\right)^2\\
\sumi (\siz)^2&
=
\left( \sum_{i=1}^{N/2}\Big(\frac{\cta}{\sqrt{2C}}(a_i+\aidag) +\sta \big(1-\frac{\aidag a_i}{C}\big)\Big)^2
+\sum_{i=N/2+1}^N \Big(\frac{\cta}{\sqrt{2C}}(a_i+\aidag) -\sta \big(1-\frac{\aidag a_i}{C}\big)
\Big)^2 \right)\nonumber\\
&\to
 \sumi \left(
\frac{\cos^2\theta}{2C}(a_i+\aidag)^2-\frac{2\sin^2\theta}{C}\aidag a_i \right)\\
-C\sumi \six&=C \sum_{i=1}^{N/2} \left(\sta \frac{a_i+\aidag}{\sqrt{2C}}
-\cta \big(1-\frac{\aidag a_i}{C}\big)\right)
-C\sum_{i=N/2+1}^{N} \left(\sta \frac{a_i+\aidag}{\sqrt{2C}}
+\cta \big(1-\frac{\aidag a_i}{C}\big)\right)\nonumber\\
&\to  \cta\sumi \aidag a_i.
\end{align}
The quadratic term of the Hamiltonian is therefore
\begin{align}
H_2&=\frac{JC\cos^2\theta}{2N}\left(\sumi (a_i+\aidag)\right)^2
-\frac{JC\lambda\cos^2\theta}{2}\sumi (a_i+\aidag)^2
+2JC\lambda \sin^2\theta \sumi \aidag a_i +\Gamma \cta\sumi \aidag a_i.
\end{align}
For diagonalization, we apply the Fourier transformation
\begin{equation}
a_j=\frac{1}{\sqrt{N}}\sumk e^{ikj}b_k,~a_j^{\dagger}=\frac{1}{\sqrt{N}}\sum_k e^{-ikj}b_k^{\dagger}.
\end{equation}
which gives
\begin{align}
&\sum_j (a_j+a_j^{\dagger})=\sqrt{N}(b_0+b_0^{\dagger})\\
&\sum_j a_j^{\dagger}a_j =\sumk b_k^{\dagger}b_k\\
&\sum_j(a_j+a_j^{\dagger})^2= \frac{1}{N}\sumk \sum_{k'} \sum_j (e^{ikj}b_k +e^{-ikj}b_k^{\dagger})
(e^{ik'j}b_{k'}+e^{-ik'j}b_{k'}^{\dagger})\nonumber\\
&=\sumk (b_k b_{-k} +b_k b_k^{\dagger}+b_k^{\dagger}b_k +b_k^{\dagger}b_{-k}^{\dagger}).
\end{align}

The Hamiltonian is rewritten as
\begin{align}
H_2&=\frac{JC\cos^2\theta}{2N} N(b_0+\b0d)^2-\frac{JC\lambda\cos^2\theta}{2}\sum_k
(b_kb_{-k}+b_k\bkd +\bkd b_k +\bkd \bmkd) \nonumber\\
&+(2JC\lambda \sin^2\theta +\Gamma\cos^2\theta)\sum_k \bkd b_k\nonumber\\
&=\left(\frac{JC\cos^2\theta}{2}-\frac{JC\lambda\cos^2\theta}{2}\right)(b_0+\b0d)^2
+(2JC\lambda\sin^2\theta +\Gamma \cos\theta)\b0d b_0 \nonumber\\
&+\sum_{k\ne 0}\left\{ -\frac{JC\lambda \cos^2\theta}{2}(b_kb_{-k}+\bkd\bmkd)
+(2JC\lambda \sin^2\theta +\Gamma \cos\theta -JC\lambda \cos^2\theta)\bkd b_k\right\}
+{\rm const.}\nonumber\\
&=\frac{JC}{2}(1-\lambda)\cos^2\theta \big(b_0^2+(\b0d)^2\big)
+\big(JC(1-\lambda)\cos^2\theta +2JC\lambda \sin^2\theta +\Gamma \cos\theta\big)\b0d b_0 \nonumber\\
&+\sum_{k\ne 0}\left\{ -\frac{JC\lambda \cos^2\theta}{2}(b_kb_{-k}+\bkd\bmkd)
+(2JC\lambda \sin^2\theta +\Gamma \cos\theta -JC\lambda \cos^2\theta)\bkd b_k\right\}\nonumber\\
&\equiv A\left\{ \b0d b_0 +\frac{JC(1-\lambda)\cos^2\theta }{2A}\big(b_0^2+(\b0d)^2\big)\right\}
+B\sum_{k\ne 0}\left\{\bkd b_k-\frac{JC\lambda \cos^2\theta}{2B}(b_k b_{-k}+\bkd \bmkd)\right\},
\label{H2bb}
\end{align}

\twocolumngrid

where
\bes
\begin{align}
A&=JC(1-\lambda)\cos^2\theta +2JC\lambda\sin^2\theta +\Gamma 
\cta\\
B&=2JC\lambda \sin^2\theta -JC\lambda \cos^2\theta +\Gamma \cta .
\end{align}
\ees
The Hamiltonian ~\eqref{H2bb}) can be diagonalized by the Bogoliubov transformation
\bes
\begin{align}
b_k&=\alpha_k \cosh\phi +\alpha_{-k}^{\dagger} \sinh \phi \\
b_k^{\dagger}&=\alpha_k^{\dagger}\cosh \phi +\alpha_{-k}\sinh \phi
\end{align}
\ees
with
\begin{equation}
\tanh 2\phi=-\lambda,
\end{equation}
with $\lambda_k$ being defined by
\begin{equation}
\bkd b_k +\frac{\lambda}{2}(b_k b_{-k}+\bkd \bmkd).
\end{equation}
We thus find
\bes
\begin{align}
&H_2=\omega_0 \alpha_0^{\dagger}\alpha_0 +\sum_{k\ne 0}\omega_1 \alpha_k^{\dagger}\alpha_k,\\
&\omega_0=A\sqrt{1-\lambda_A^2},\quad \omega_1=B\sqrt{1-\lambda_B^2}
\end{align}
\ees
with
\begin{equation}
\lambda_A=\frac{JC(1-\lambda)\cos^2\theta}{A}, \quad \lambda_B=-\frac{JC\lambda \cos^2\theta}{B}.
\end{equation}
The energy gap is therefore the smaller of $\omega_0$ and $\omega_1$, the latter
being highly degenerate.

When $\theta =0~(\Gamma >\Gamma_{\rm c}=2JC\lambda)$,
\begin{align}
\omega_0^2&=A^2-A^2\lambda_A^2 =(JC(1-\lambda)+\Gamma)^2-J^2C^2(1-\lambda)^2\notag \\
&=\Gamma^2 +2JC(1-\lambda)\Gamma >0 ~(\lambda<1)\\
\omega_1^2&=B^2-B^2\lambda_B^2 =(\Gamma-JC\lambda)^2-(JC\lambda)^2\notag \\
&=\Gamma (\Gamma -2JC\lambda)\ .
\end{align}
Therefore, $\omega_0$ does not vanish whereas $\omega_1$ behaves around $\Gamma =\Gamma_{\rm c}$ as
\begin{equation}
\omega_1 =\sqrt{\Gamma (\Gamma -2JC\lambda)}\approx \sqrt{2JC\lambda}\sqrt{\Gamma-\Gamma_{\rm c}}.
\label{og1a}
\end{equation}
This is how the gap closes in the thermodynamic limit. Similarly, for $0\le \theta \le \pi /2~(0\le \Gamma <\Gamma_{\rm c}=2JC\lambda)$, using
\begin{equation}
\cos\theta =\frac{\Gamma}{2JC\lambda},
\end{equation}
we have
\bes
\begin{align}
A&=JC(1-\lambda)\frac{\Gamma^2}{(2JC\lambda)^2}+2JC\lambda \left(1-\frac{\Gamma^2}{(2JC\lambda)^2}\right)\notag \\
&+\frac{\Gamma^2}{2JC\lambda}
=\frac{1-\lambda}{4JC\lambda^2}\Gamma^2 +2JC\lambda,\\
A\lambda_A&=JC(1-\lambda)\frac{\Gamma^2}{(2JC\lambda)^2}=\frac{1-\lambda}{4JC\lambda^2}\Gamma^2.
\end{align}
\ees
Hence,
\begin{align}
A^2-A^2\lambda_A^2&=\left(\frac{1-\lambda}{4JC\lambda^2}\Gamma^2 +2JC\lambda\right)^2
-\left(\frac{1-\lambda}{4JC\lambda^2}\Gamma^2\right)^2\nonumber\\
&=(2JC\lambda)^2 +2JC\lambda \frac{1-\lambda}{2JC\lambda^2}\Gamma^2 >0.
\end{align}
This means that the zero-mode does not vanish, as was the case for $\Gamma>\Gamma_{\rm c}$. The other mode has
\bes
\begin{align}
B&=2JC\lambda \left(1-\frac{\Gamma^2}{(2JC\lambda)^2}\right)-JC\lambda \frac{\Gamma^2}{(2JC\lambda)^2}\notag \\
&+\frac{\Gamma^2}{2JC\lambda}=2JC\lambda -\frac{\Gamma}{4JC\lambda}\\
B\lambda_B&=-JC\lambda \frac{\Gamma^2}{(2JC\lambda)^2}=-\frac{\Gamma^2}{4JC\lambda}\\
B^2-B^2\lambda_B^2&=\left( 2JC\lambda -\frac{\Gamma^2}{4JC\lambda}+\frac{\Gamma^2}{4JC\lambda}\right)\times\notag \\
&\quad\left( 2JC\lambda -\frac{\Gamma^2}{4JC\lambda}-\frac{\Gamma^2}{4JC\lambda}\right)\nonumber\\
&=2JC\lambda \left(2JC\lambda -\frac{\Gamma^2}{2JC\lambda}\right)\notag\\
&=(2JC\lambda -\Gamma)(2JC\lambda +\Gamma)\ .
\end{align}
\ees
Then for $\Gamma \approx \Gamma_{\rm c}$,
\begin{equation}
\omega_1 \approx \sqrt{4JC\lambda}\sqrt{\Gamma_{\rm c}-\Gamma}.
\label{og1b}
\end{equation}
The results~\eqref{og1a} and~\eqref{og1b} show that the gap closes with exponent
$1/2(=\nu z)$ near the critical point.  The exponent that expresses the rate of gap closing
at the critical point for finite systems is $z$, with $\Delta \sim L^{-z}$ (though it is
unclear what $L$ is in the present infinite-range model).
The critical exponent $\nu z$ is independent of the non-universal parameter $C$, as expected,
and the same is expected for $z$.
Therefore, we must accept that nesting does not qualitatively improve the rate of gap closing
($C$-independence of $z$).
Nevertheless, the coefficients in Eqs.~\eqref{og1a} and ~\eqref{og1b} increase
proportionally to $\sqrt{C}$, and hence the gap becomes {\em quantitatively} large
for large $C$. 

One of the strengths of this method is that it applies to arbitrary lattices,
not just the mean-field model, including Chimera-native problems as long as $C\gg 1$. 
In such cases, the final gap spectrum $\omega_k$ will have a $k$-dependence.
The present results Eqs.~\eqref{og1a} and~\eqref{og1b} apply directly only to the case $C\gg 1$.
Nevertheless, the exponent $1/2$ appearing in these expressions should not depend on $C$ and therefore
would be valid for smaller $C$ such as $C=4$.  The same would apply to other critical exponents.
The coefficients in these equations, $\sqrt{2JC\lambda}$ and $\sqrt{4JC\lambda}$,
 would be the leading terms of asymptotic expansions of the relevant coefficient $g(C)$
in the limit of $C\gg 1$. In other words,
it is expected that the gap closes as $g(C)\sqrt{|\Gamma-\Gamma_{\rm c}|}$
for general values of $C$ with $g(C)$ having the above asymptotic forms for $C\gg 1$.

\bibliography{refs}


\end{document}